\documentclass[fleqn,12pt]{wlscirep}
\usepackage{wrapfig, framed, caption}
\usepackage[utf8]{inputenc}
\usepackage{heuristica}
\usepackage{graphicx}
\usepackage{wrapfig, framed, caption}
\usepackage{tcolorbox}
\usepackage{dirtytalk}
\usepackage{xr}
\usepackage{setspace}
\usepackage[normalem]{ulem}
\useunder{\uline}{\ulined}{}
\usepackage{xparse}
\newsavebox{\fminipagebox}
\NewDocumentEnvironment{fminipage}{m O{\fboxsep}}
 {\par\kern#2\noindent\begin{lrbox}{\fminipagebox}
  \begin{minipage}{#1}\ignorespaces}
 {\end{minipage}\end{lrbox}%
  \makebox[#1]{%
    \kern\dimexpr-\fboxsep-\fboxrule\relax
    \fbox{\usebox{\fminipagebox}}%
    \kern\dimexpr-\fboxsep-\fboxrule\relax
  }\par\kern#2
 }

\title{Taking Census of Physics}

\author[1]{Federico Battiston}
\author[1]{Federico Musciotto}
\author[2,3]{Dashun Wang}
\author[1,4,5]{Albert-L\'aszl\'o Barab\'asi}
\author[1,4,6,7]{Michael Szell}
\author[1,8,4,6,9*]{Roberta Sinatra}

\affil[1]{Department of Network and Data Science, Central European University, Budapest, 1051, Hungary}
\affil[2]{Kellogg School of Management, Northwestern University, Evanston, IL 60208, USA}
\affil[3]{Northwestern Institute on Complex Systems, Northwestern University, Evanston, IL 60208, USA}
\affil[4]{Network Science Institute, Northeastern University, Boston, MA 02115, USA}
\affil[5]{Center for Cancer Systems Biology, Dana-Farber Cancer Institute, Boston, MA 02115, USA}
\affil[6]{Complexity Science Hub Vienna, Vienna, 1080, Austria}
\affil[7]{MTA KRTK Agglomeration and Social Networks Lendulet Research Group, Centre for Economic and Regional Studies, Hungarian Academy of Sciences, Budapest, 1094, Hungary}
\affil[8]{Department of Mathematics, Central European University, Budapest, 1051, Hungary}
\affil[9]{ISI Foundation, Torino, 10126, Italy}

\affil[*]{robertasinatra@gmail.com}

\begin{document}

\maketitle

%
%
%
%
%
%

\thispagestyle{empty}
There was a time when polymaths like Galileo knew all the physics that was there to be known. Over the centuries, however, the body of knowledge spanned by physics exploded, encompassing topics as diverse as gravitational waves, graphene, or network science. As physics expanded in breadth and depth, physicists were forced to specialise \cite{jones2009burden}, segmenting researchers into their narrow, specialised communities.
How many physicists work in each subfield of physics today and how does each subdiscipline evolve? In which subfield are physicists ``born'' into and where do they migrate, if at all? Here we take an intellectual census of physicists, their activities and career trajectories, helping us understand the evolution of the field and gaining quantitative insights about several fundamental scientific processes, from resource allocation to the exchange of knowledge. Advances in this direction were limited by the challenge in answering two fundamental questions: 1) Who can be counted as a physicist? 2) How do we survey their activities? The recent availability of large datasets of scientific publications finally offers opportunities to tackle these questions by exploring the production patterns of the scientific population~\cite{clauset2017data, Fortunato2018}. Indeed, the close to complete publication records of all physicists allow us to reconstruct their subfields of study and career changes, offering quantitative footprints not just for the field of physics, but its intimate relation with the broader scientific community~\cite{deville2014career, Sinatra2015}.

Combining large-scale data on physics publications and citations with recent data and network science techniques, here we ask: What are the impact and productivity differences between subfields? As a physics student choosing my future specialty, how do I know which subfields are growing? As a funding agency, how do I compare early-career physicists from different subfields? As a journal editor, how many papers should I expect from each subfield and how do I compare their impact? 

\subsection*{A census of physics subfields}

To offer a data-driven answer to these questions~\cite{clauset2017data, Fortunato2018}, we identify the relevant physics papers and citations within Web of Science (WoS). We start by selecting $\sim$3.2 million physics papers, published in 294 physics journals indexed by WoS. This core represents, however, only a fraction of all physics papers ~\cite{Sinatra2015, DevilleThesis}, missing for example those published in interdisciplinary journals like \textit{Nature} or \textit{Science}, or papers published in journals of other disciplines but that are of direct relevance for the physics community. To map out the complete physics literature we then set to detect physics papers by virtue of their patterns of citations among the other $\sim$47 million papers in WoS. A paper is a potential physics publication if its references and citations to the core physics literature are significantly higher than in a null model in which each paper's citations are assigned
randomly, regardless of a paper's journal or research area. We identified $\sim$4.5 million papers whose patterns of citations and references are indistinguishable from papers in physics journals (SI Section S1), obtaining overall a dataset of $\sim$7.7 million publications of interest to the physics community.

We characterise further this physics corpus by classifying each paper into nine major subfields, according to the Physics and Astronomy Classification Scheme (PACS) used by the American Physical Society~\cite{PacsAPS} between 1985 and 2015. We use this dataset to reconstruct the publication profile of 135,877 physicists with a persistent productivity between 1985 and 2015. See Box~1 and SI Section S3 for more details on the dataset curation and validation. 

\begin{figure}[ht]
\centering
\includegraphics[width=0.8\textwidth]{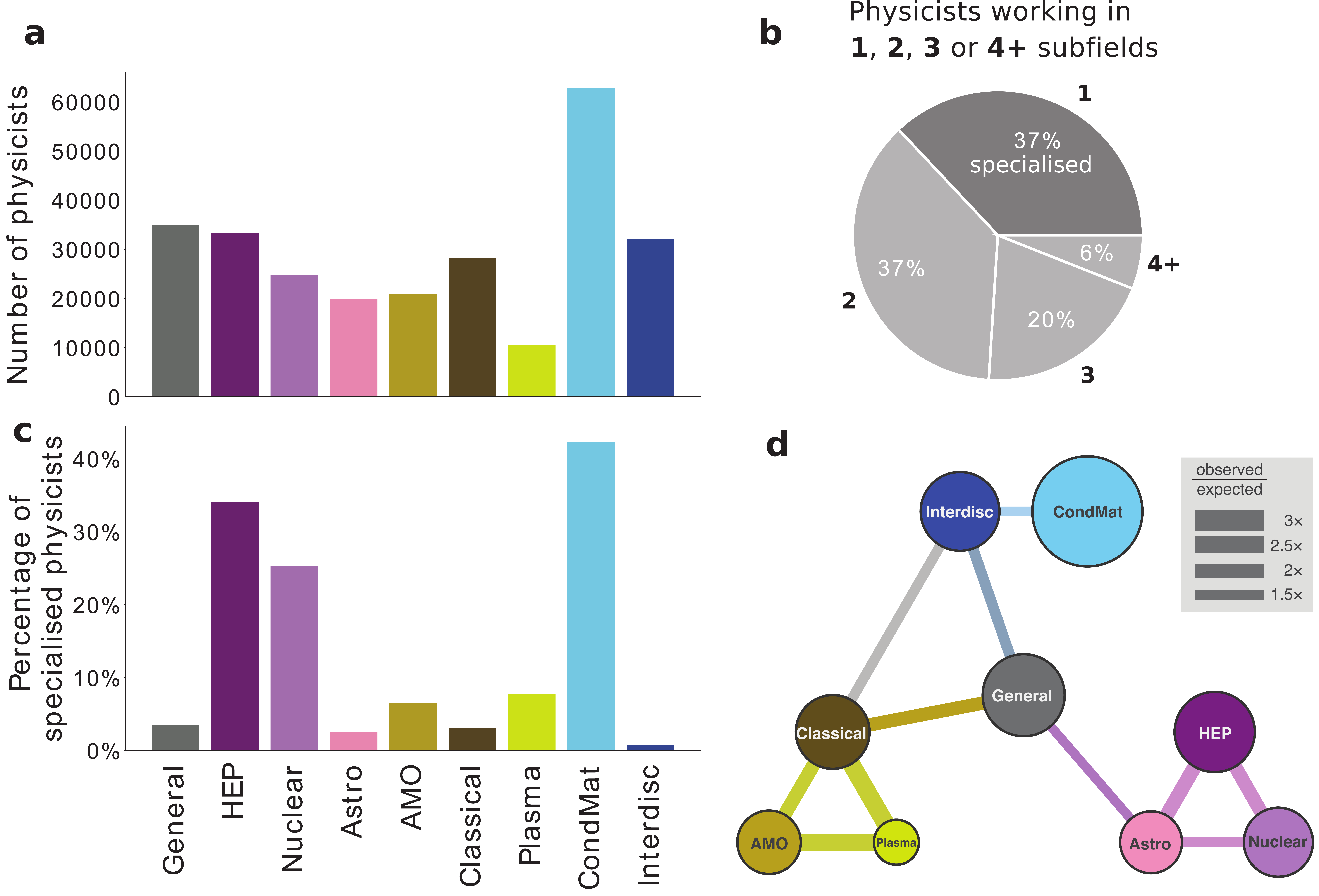}
\caption{\label{fig:fields}
\textbf{Taking census of physics subfields.} \textbf{a}, Number of physicists per subfield. \textbf{b}, Percentage of physicists working in 1, 2, 3, or 4+ subfields. We call the 37\% of physicists who work in only one subfield \emph{specialised}. \textbf{c}, Fraction of specialised physicists per subfield. Most subfields except for {\sf HEP}, {\sf Nuclear} and {\sf CondMat} have a negligible fraction of specialised physicists.
\textbf{d}, The network of co-activity of individual physicists shows the nontrivial connection between subfields. Node size is proportional to number of physicists in the subfield, link width is proportional to the overlap between subfields, quantified with the ratio between measured number of physicists working on the two subfields and 
expected number based on a randomised null model.}
\end{figure}

The first step in developing a census is to count the number of physicists working in each subfield. Such counting is, however, not straightforward, as physicists may contribute to publications in different subfields. We therefore associate each physicist with a primary subfield if the number of her publications in the subfield is higher, in a statistically significant manner, than expected for a typical physicist (Box~1 and SI Section S4). The obtained subfield demographics offer us a first summary statistic (Fig.~\ref{fig:fields}a): we find that the largest subfield is {\sf CondMat} (condensed matter physics) with more than $62,000$ physicists, capturing 46\% of the entire physicist population. It is followed by {\sf General} ($34,000$), {\sf HEP} (high energy physics, $33,000$), Interdisc (Interdisciplinary physics, $32,000$), {\sf Classical} ($28,000$), {\sf Nuclear} ($24,000$), {\sf AMO} (Atomic and molecular physics, $20,000$) and {\sf Astro} physics ($19,000$). {\sf Plasma} is the smallest subfield of physics, with less than $11,000$ researchers.

Given the highly specialised nature of the physics subfields, one might suspect that most physicists work in a single subfield. Yet, we find that highly specialised physicists are the exception rather than the rule: The majority of physicists (63\%) are active in two or more subfields (Fig.~\ref{fig:fields}b). This prompts us to ask: Which subfields have particularly low or high rates of specialisation? The differences between subfields are striking, defining two different groups (Fig.~\ref{fig:fields}c): six subfields have less than $10\%$ specialised physicists. Among these subfields, {\sf Interdisc} has less than 1\% of specialised physicists, in line with the expectation that interdisciplinary physicists bridge multiple subfields. In contrast, the percentage of specialised physicists in {\sf CondMat}, {\sf HEP} and {\sf Nuclear} is 42\%, 34\% and 25\% respectively, at least an order of magnitude larger than in the other group of subfields. What drives the different levels of specialisation between subfields?

A physicist working on two or more subfields combines the collective know-how of these fields, a process deemed essential for novel discoveries in science \cite{dyson2009bf,Uzzi2013,Foster2015tradition}. To understand which of the physics subfields cross-pollinate most significantly, we calculate the co-activities of individual physicists between each pair of subfields. Co-activities are defined by weighted links between subfields, where the weights measure the observed versus expected co-activities based on a randomised null model (SI Section S6). Starting with the highest weighted links, we plot the minimum number of links needed to have a connected network of subfields (Fig.~\ref{fig:fields}d). The network reveals a non-trivial co-activity structure, clustering all physics subfields into three broader areas, 1) {\sf Interdisc} and {\sf CondMat}, 2) {\sf Classical}, {\sf AMO}, and {\sf Plasma}, 3) {\sf HEP}, {\sf Astro}, and {\sf Nuclear}, all held together by {\sf General}. This research space captures the intellectual affinities between subfields, facilitating movements between close subfields, while limiting cross-pollination between distant ones like {\sf Interdisc} and {\sf Nuclear}\cite{guevara2016rsu}. For example, the diversity of topics within {\sf CondMat} and {\sf Classical} and their adaptable approaches, like statistical mechanics applied to multiple systems composed of large numbers of entities, makes it easier for those working in these subfields to take their tools to different disciplines.  In contrast, more specialised subfields like {\sf HEP} or {\sf Nuclear} require their members to acquire familiarity with large-scale, long-term projects. While scientists working in such fields may have deep knowledge and expertise on the subject they specialise in, they face a greater burden that limits their ability to explore other areas. The observed network is similar to the citation network\cite{Sinatra2015} between subfields, showing that the flow of knowledge is captured through multiple metrics, both by paper citations and by the activities of individual physicists\cite{guevara2016rsu}.

\subsection*{Birth, growth, and migration}
\begin{figure}[ht]
\centering
\includegraphics[width=0.8\textwidth]{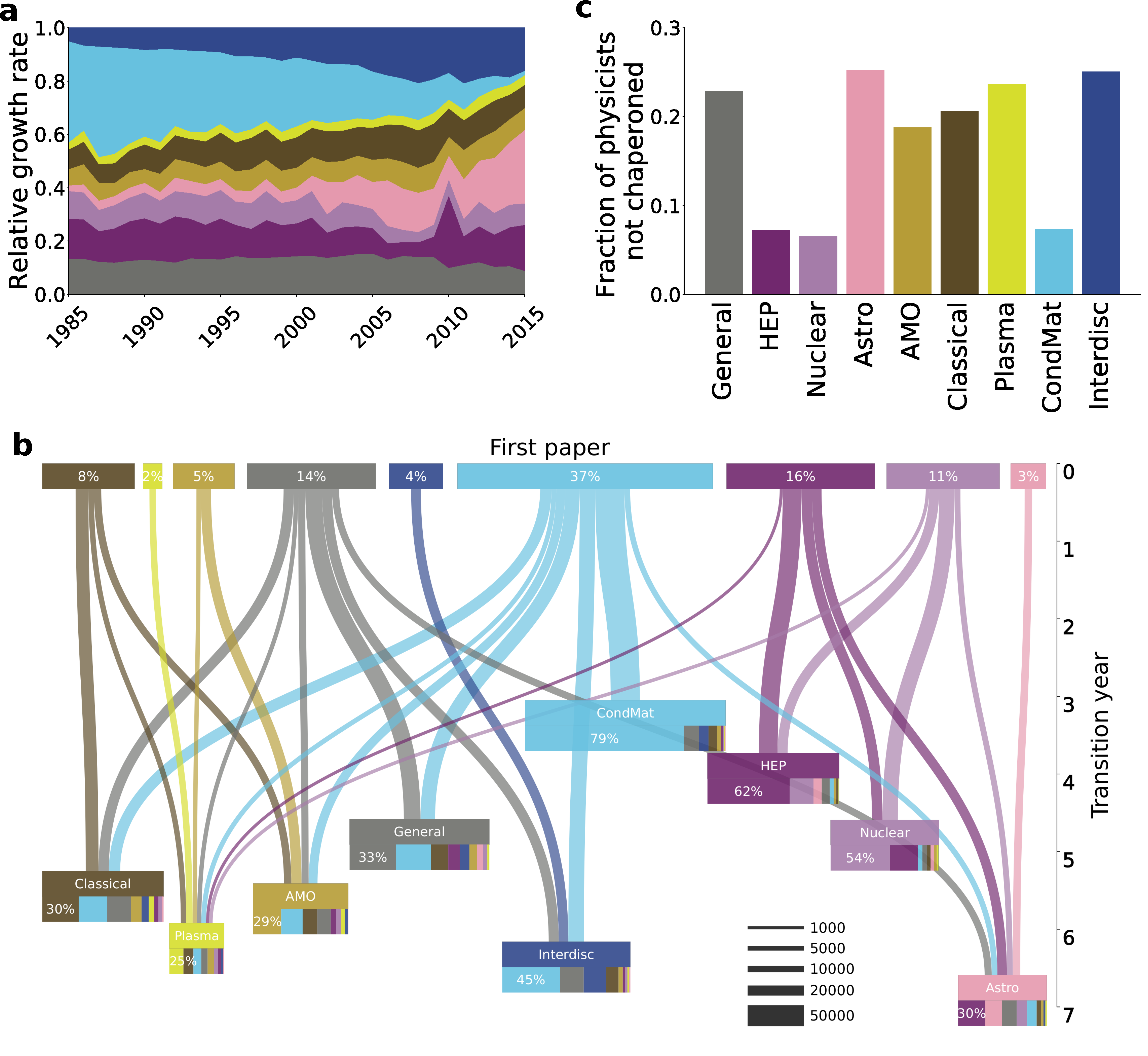}
\caption{\textbf{Evolution of physics subfields and careers.} \textbf{a}, Relative growth rate, defined as yearly fraction of physicists who published their first paper in a new subfield. {\sf Interdisc} and {\sf Astro} grow, {\sf CondMat} shrinks considerably. {\sf HEP} displays a spike in 2010 that can be attributed to large-scale collaborations like ATLAS and CMS (SI Section S7). Relative growth rate is less reliable after 2010 due to early-career physicists accumulating publications at different rates in each subfield, resulting in reaching the 5 publications threshold at different times and distorting the proportion of physicists in favor of more productive and non-specialised subfields. \textbf{b}, Flow diagram of career transitions. The sizes of rectangles on the top are proportional to the number of career first publications in each given subfield. The rectangles at the bottom are proportional to the number of physicists in each subfield who did not start their career by publishing in the area -- for example {\sf Astro} and {\sf AMO} have roughly the same number of physicists although {\sf Astro} starts with 3\%, while AMO with 5\%. The distance from the top reflects the average time at which a career transition towards a subfield occurs. Flows are proportional to the number of physicists who first published in a subfield different from the one in which they worked previously. Only significant flows, i.e. those that are larger than expected in the null model, are shown. The percentages on the bottom rectangles report the contribution of the subfield that is contributing most. \textbf{c}, Fraction of not chaperoned physicists in each subfield. A large majority of physicists starting in {\sf HEP}, {\sf Nuclear}, or {\sf CondMat} co-author their first paper with physicists who have already published in the subfield. Other subfields have a much higher fraction of physicists who are not chaperoned in.}
\label{fig:transition}
\end{figure}

Why are there so considerable differences in specialised physicists between similarly sized subfields, like {\sf Nuclear} and {\sf Interdisc} (Fig.~\ref{fig:fields}a,b)? 
To understand this heterogeneity, we first assess the relative growth rate of each subfield over time, measuring the fraction of physicists entering a subfield every year (Fig.~\ref{fig:transition}a). We find that the growth rates of {\sf Interdisc} and {\sf Astro} increased from a few percent in 1985 to over 20\% and 27\% respectively after 2010, substantially reshaping the physics landscape in recent years. An opposite trend characterises {\sf CondMat}: while it had the largest share of new physicists in 1985, its share dramatically decreased over time, falling below 5\% after 2010. {\sf HEP} also displayed a receding trend just before 2010, but the spur of new research connected to the activity of the Large Hadron Collider in Geneva injected new forces into the field. In particular, {\sf HEP}'s sharp peak in 2010 can be attributed to the first ATLAS and CMS publications~\cite{atlas} (SI Section S7).

Figure~\ref{fig:transition}a mixes together physicists who start their careers in a particular subfield with those who make career transitions to other subfields. There are remarkable examples of physicists who never changed their subfield, like Klaus von Klitzing, whose first publication was in {\sf CondMat}, and contributed over 500 papers to the subfield, earning him the Nobel Prize in 1985 for the discovery of the quantised Hall effect. In contrast, Rainer Weiss, best known for inventing the laser interferometric technique at the heart of LIGO, which earned him the Nobel Prize in 2017, published his first paper on an unrelated topic in {\sf AMO}, ``Magnetic Moments and Hyperfine-Structure Anomalies of $Cs_{133}$, $Cs_{135}$ and $Cs_{137}$''. To distinguish such different careers, we next systematically explore career transitions within physics~\cite{Jia2017}, asking: Where are physicists ``born'', and how do they ``migrate'' between subfields? When do these transitions typically occur? 

Figure~\ref{fig:transition}b shows how many physicists began their careers in each subfield (top rectangles). Remarkably, $64\%$ of the physicists began their careers by publishing in either {\sf CondMat}, {\sf HEP}, or {\sf Nuclear} ($37\%$ of all physicists start out in {\sf CondMat}). These three subfields capture ``curricular'' physics topics, the natural ending points of many undergraduate courses, hence the typical starting point of research careers. {\sf General}, covering topics of interests to a wide set of physicists, accounts for $14\%$ of first publications. In contrast, only $4\%$ of physicists started publishing in {\sf Interdisc}, and as low as $3\%$ began in {\sf Astro}. As {\sf Interdisc} integrates other disciplines, it might be difficult to start out as an {\sf Interdisc} physicist; the low percentage of {\sf Astro} starts may be rooted in the fact that traditionally it has not been a ``curricular'' subfield.


\begin{center}
\linespread{1.5}
\centering
\begin{tcolorbox}[before=\par\smallskip\centering,after=\par,sharp corners, colback=red!2!green!8!blue!10, colframe=red!2!green!8!blue!20,coltitle=black, title={\bf Box 1},width=1.\textwidth, on line,
  grow to left by=0.5cm,
  grow to right by=0.5cm]
{\footnotesize

{\bf Identifying subfields}\\
  We classify papers into 9 subfields, based on the 1-digit Physics and Astronomy Classification Scheme (PACS) by the American Physical Society (APS):
  \begin{itemize}
  \setlength\itemsep{0em}
	\item{\sf General}: Mathematical Methods, Quantum Mechanics, Relativity, Nonlinear Dynamics, Metrology
	\item {\sf HEP:} The Physics of Elementary Particles and Fields
	\item {\sf Nuclear:} Nuclear Structure and Reactions
	\item {\sf AMO:} Atomic and Molecular Physics
	\item {\sf Classical:} Electromagnetism, Optics, Acoustics, Heat Transfer, Classical Mechanics, and Fluid Dynamics
	\item {\sf  Plasma:} Physics of Gases, Plasmas, and Electric Discharges
	\item {\sf CondMat:} Structural, Mechanical, and Thermal Properties; Electronic Structure, Electrical, Magnetic, and Optical Properties
	\item {\sf Interdisc:} Interdisciplinary Physics and Related Areas of Science and Technology
	\item {\sf Astro:} Astrophysics, Astronomy, and Geophysics
\end{itemize}
  
PACS were consistently used in papers published in APS journals between 1985 and 2015 (SI Section S2). Using an algorithm that evaluates the patterns of citations and references between papers, we propagate subfield labels from APS papers to other papers: if the fraction of references and citations between a given paper and papers in a particular subfield is larger than expected by the null model, the paper is assigned to that subfield. A paper may be assigned to multiple subfields, in line with APS papers reporting multiple PACS. In panel a) we show an example of an unclassified paper which references in {\sf CondMat}, {\sf Plasma} and {\sf Astro}, and which is cited by {\sf CondMat}, {\sf Astro} and another publication still lacking a PACS. The publication is first assigned to {\sf CondMat} and then to {\sf Astro}, but not to {\sf Plasma}, as it lacks statistical significant links to the subfield. The algorithm is run iteratively until convergence for each subfield, helping us associate at least one subfield to 1,137,670 papers (SI Section S3).\\[-6pt] 

{\bf Assigning physicists to subfields}\\
We analyse all careers with at least $5$ labeled papers between 1985 and 2015, capturing the careers of 135,877 physicists. We consider a physicist working in a subfield if her share of publications in the subfield is higher than that of the average physicist. The statistical criterion ~\cite{balassa1965} we used, guarantees that each scientist is assigned to at least one subfield, and takes into account the different sizes of subfields. As an example, we show the result of the criterion applied to the career of Stephen Hawking in panel b). In the physics dataset Hawking has 124 papers associated to different subfields. Of these subfields, only {\sf General} (95 papers) and {\sf Astro} (77 papers) are assigned to the physicist through the statistical criterion, whereas {\sf HEP} (23 papers) and {\sf Classical} (1 paper) are not statistically significant, which is consistent with Hawking being known as a theoretical physicist and cosmologist.\\[-6pt] 

For validation and further methods see SI Sections S3, S4, S5.
}
\end{tcolorbox}
\end{center}

 \linespread{1.5}
\begin{tcolorbox}[before=\par\smallskip\centering,after=\par,sharp corners, colback=red!2!green!8!blue!10, colframe=red!2!green!8!blue!20,width=1.\textwidth,
grow to left by=0.5cm,
  grow to right by=0.5cm]


\centering
\includegraphics[width=1.0\textwidth]{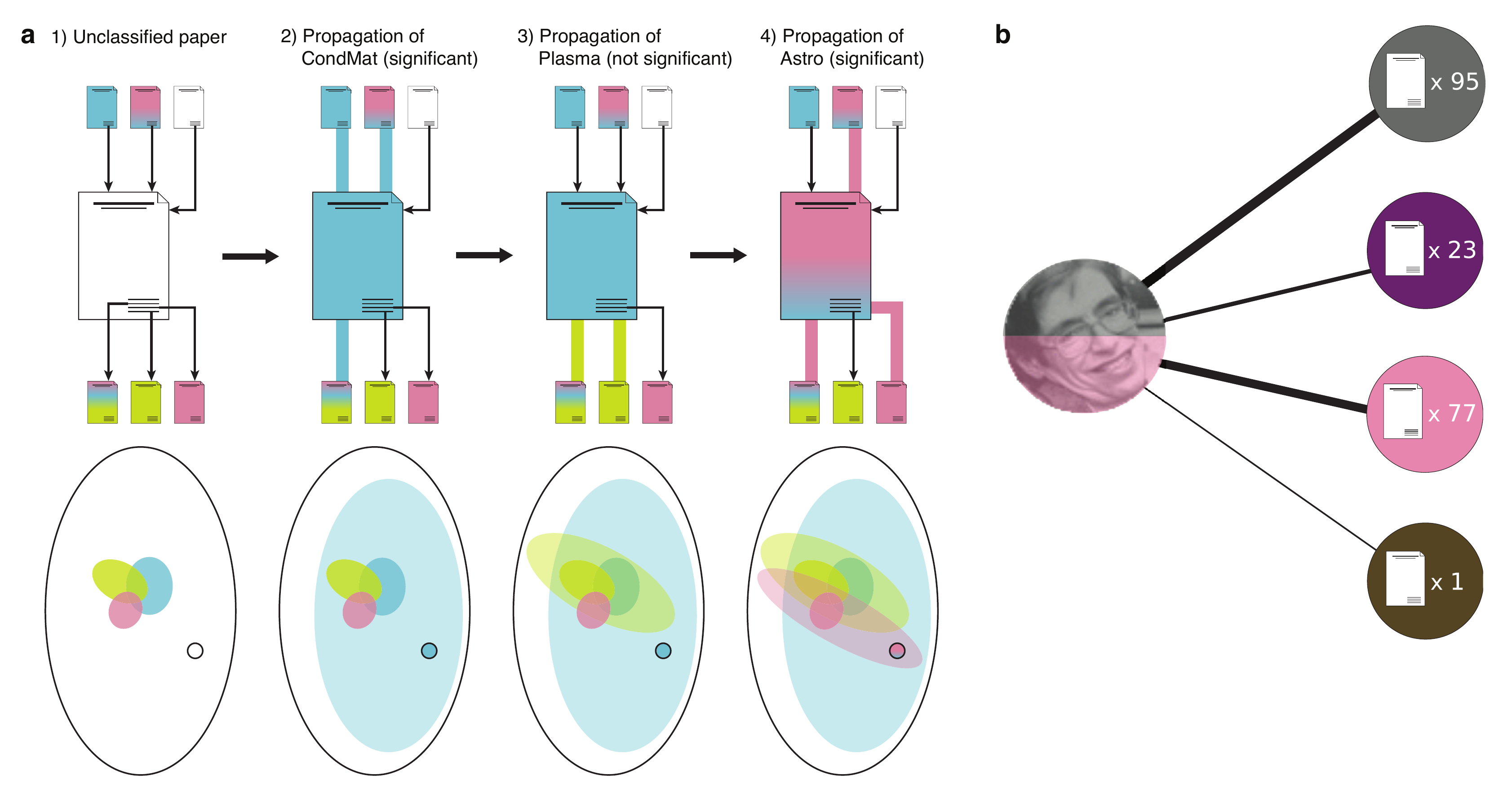}

\end{tcolorbox}



The links of Fig.~\ref{fig:transition}b capture the significant flows between subfields, linking the subfield where a physicist published her first paper, to the subfields that best characterised her later careers (SI Section S6).
 This diagram indicates that {\sf CondMat} is the starting point for many physicists who later specialised in {\sf Interdisc}, {\sf Classical}, and {\sf General}. {\sf HEP} and {\sf Nuclear} tend to swap researchers while feeding talents into {\sf Astro}, a pattern that may be rooted in the fact that all three subfields study radiation or nuclear and subnuclear processes. We find that most {\sf Interdisc} physicists did not start their career there, but migrated from {\sf CondMat} and {\sf General}, consistent with the hypothesis that one needs to acquire expertise in at least two fields before being able to bring them together. Finally, {\sf Plasma} and {\sf Astro} welcome physicists with many different backgrounds, but rarely feed into other subfields. The diversity of the incoming flows to {\sf Plasma} and {\sf Astro} suggests their accessibility to physicists with many different backgrounds.

We also measure the average time it takes to transition to a different subfield, captured by the vertical axis of Fig.~\ref{fig:transition}b. Once again, {\sf HEP}, {\sf Nuclear} and {\sf CondMat} top the list: physicists who did not start their career in these subfields tend to transition towards them the earliest, typically by the third or fourth year of their research career. The opposite trend was observed for {\sf Interdisc} and {\sf Astro}, which not only have the highest transition rates among subfields, but are also characterised by the longest time to transition. Indeed, on average a physicist publishes her first paper on these two topics $6$ to $7$ years into her career, roughly double the transition time towards {\sf HEP}, {\sf Nuclear} and {\sf CondMat}. {\sf Interdisc} displays a late switch, consistent with the hypothesis that it takes time to gather expertise in multiple fields. Similarly, physicists tend to switch to {\sf Astro} typically after a relatively long experience in {\sf HEP}.

The flow diagram of Fig.~\ref{fig:transition}b helps us better understand the research space captured by Fig.~\ref{fig:fields}d. For instance, in the bottom right triple, {\sf HEP} plays the leading role in producing physicists who transition to its tightly connected subfields, {\sf Nuclear} and {\sf Astro}. In the top two nodes of the network, {\sf CondMat} is the main force feeding {\sf Interdisc}. The observed widespread career transitions may reflect potential benefits to the whole field, cross-pollinating one physics community with ideas and methods developed by a different subfield~\cite{dyson2009bf,Uzzi2013}.

\subsection*{The role of chaperones}
The future prosperity of young scholars has often been linked to access to valuable mentorship at the early stages of a scientific career~\cite{Crosta2005, Malgrem2010, Chariker2016}. For example, a surprising fraction of Nobel laureates had a mentor-mentee or a co-authorship relation with another Nobel laureate \cite{zuckerman1967nobel, ma2018scientific}, and scientists who co-author early with an established scientist are more likely to have higher impact and higher chances to publish as lead author than other scientists \cite{Sekara2018}. Taken together, a senior scientist who acts as ``chaperone'' during a scientist's early career might foster the acquisition of skills, passing on experience and knowledge necessary for high achievements later in a career.

To quantify the chaperone effect, we measure how many physicists co-author their first paper in a subfield with a physicist who has published in that subfield before \cite{Sekara2018}. We find that the chaperone effect is particularly strong for {\sf HEP}, {\sf Nuclear} and {\sf CondMat}, where over $90\%$ of physicists wrote their first paper with someone who published before in the same subfield (Fig.~\ref{fig:transition}c and SI Section S8). This large share of chaperoned physicists could have several reasons, like the documented  high number of physicists starting their career in these three subfields, or the need to access large facilities, which require early-career physicists to collaborate with established scientists. Note that the typical large co-authorships patterns of {\sf HEP} can not explain the magnitude of the chaperone effect characterising this subfield (SI Section S8).

Other subfields have a lower fraction of chaperoned physicists, especially {\sf Interdisc} and {\sf Astro}. These subfields are often explored by more senior physicists who received mentorship at a previous stage of their careers in a different subfield and often decide to explore the new area without close supervision (26\% of physicists are not chaperoned in {\sf Interdisc} and {\sf Astro}, Fig.~\ref{fig:transition}c). 
On top of this, 
applications of computational physics, like computational biophysics or complex systems, classified as {\sf Interdisc}, require lower financial resources compared to experimental research and could also play a significant role in explaining the low chaperone effect \cite{szell2015research}.
Taken together, the chaperone effect is strong in physics, with an average rate of 82\% chaperoned physicists across subfields. The effect signals a research culture where physicists often get introduced to their future research area by senior colleagues in a collaborative setting, in contrast with disciplines like mathematics, where the majority of scientists start their career with publishing solo-author papers\cite{Sekara2018}.

\subsection*{Productivity, impact, and team size across subfields}
Productivity and impact, capturing the number of papers published and citations received by a physicist, are frequently used metrics in the assessment of scientific careers~\cite{Sinatra2016,Liu2018}. These quantities have implications for decisions and policies involving predicting, nurturing, and funding early career scientists. Yet, the proper interpretation of these metrics must account for the highly heterogeneous productivity and citation patterns characterising different subfields ~\cite{Radicchi2008} and for different team sizes~\cite{Pavlidis2014}, both of which vary in time. 

Team size, i.e. the number of coauthors per paper, has been increasing steadily over the past decades in all fields, capturing an increasing collaboration in science~\cite{Wuchty2007}. Are there particular differences in collaborative patterns in the different physics subfields, and what are their implications on productivity and impact? To answer this question, we assess the diversity and evolution of collaboration, productivity, and citation standards in the different subfields of physics. First, the tendency of scientists to work in increasingly large teams has been particularly pronounced in {\sf HEP} (especially after 2005), {\sf Nuclear} (especially after 2010) and {\sf Astro} (especially after 2000) (Fig.~\ref{fig:impact}a). The observed explosive growth in these three subfield is partly rooted in large-scale projects like ATLAS (SI Section S7). They also result in an increased productivity: as physicists were involved in more and larger teams, the average number of papers they published each year increased by a factor of 10 for {\sf HEP} and by a factor of 2 for {\sf Nuclear} and {\sf Astro} from 1985 to 2015 (Fig.~\ref{fig:impact}b). 
However, for the other six subfields productivity has stayed constant over 30 years, and for all subfields productivity has increased at a slower rate than team sizes.
These different rates of increase 
explain why fractional productivity, i.e. the ratio between the number of papers and the average team size, decreased across all subfields (Fig.~\ref{fig:impact}c). The effect is the strongest in {\sf HEP}, {\sf Nuclear}, and {\sf Astro}, where team size grew disproportionately. It is worth noting that in these subfields authors are usually ordered alphabetically due to the large average team size, making the assessment of credits for single authors more problematic~\cite{Shen2014}.
Taken together, we find that the amount of knowledge produced per capita decreases in all subfields despite the increase in the total number of physicists and physics papers.

Given the explosive increase in both team size and the number of papers per physicists in {\sf HEP}, do {\sf HEP} physicists today have more or less impact than they had decades earlier? To answer this question we measured the average impact in number of citations after 5 years (Fig.~\ref{fig:impact}d) and the fractional impact (ratio between number of citations and average team size, Fig.~\ref{fig:impact}e) per physicist per subfield. Interestingly, the average impact of {\sf HEP} shows a growth of comparable magnitude as the growth in average productivity, leading to an unchanged fractional impact. In other words, large-scale projects like ATLAS produce papers that generate a large number of citations, compensating for the massive numbers of co-authors (hundreds or more). 

Given some of the large productivity differences between different subfields, we also expect differences in impact~\cite{Lehmann2006, Lehmann2008, Hicks2015, Waltman2016}, measured in terms of cumulative citations over a career. For instance, how much impact does it take to be a scientific leader in {\sf HEP} and how is that different in {\sf CondMat}? In Fig.~\ref{fig:impact}f and Fig.~\ref{fig:impact}g we show the total number of papers and citations acquired over an average career by the top 5\% of physicists in each subfield (in terms of productivity). In both terms, {\sf HEP} is by far the most rewarding subfield, whose top scientists coauthor 169 papers and accumulate over 7,000 citations. In contrast, top {\sf Interdisc} physicists coauthor only 18 papers with less than 1,000 citations. The large discrepancy is not explained by paper citation rates~\cite{lillquist2010,Radicchi2011}, which are roughly constant across subfields (SI Section S9), but by the high or low number of papers per author in the respective subfield (Fig.~\ref{fig:impact}b). As a consequence, when physicists with different specialties compete for positions or grants, caution is needed in comparing their profiles using metrics based on citations or productivity, as subfield-dependent differences appear from the very beginning of a career.

What about the rate of top papers in the different subfields? We selected the top 1\% of all physics papers (in terms of citations) and assessed into which subfield they fall (Fig.~\ref{fig:impact}h). The majority falls into {\sf CondMat}, {\sf General} and {\sf HEP}, however, this result is trivial as these fields produce the most papers. To unveil the significant effects we measured the surplus between this top 1\% distribution and the distribution of subfields of all physics papers. As Fig.~\ref{fig:impact}i shows, {\sf Interdisc} papers are 40\% more likely to be in the top 1\% than expected, while {\sf Nuclear} and {\sf Plasma} papers are 40\% less likely to be found in the top 1\%. The high rate of {\sf Interdisc} among the top cited papers might be partially explained by the finding that papers which are 15\% novel and 85\% conventional often have high impact \cite{Uzzi2013}. {\sf Interdisc} is more likely to achieve this balance, since interdisciplinary research must be novel and, at the same time, must adhere to established principles. Another explanation is that {\sf Interdisc} is more likely to initiate new topics or emerging subfields. Papers that do open such new avenues are known to acquire a high number of citations as they become milestones, cited by subsequent papers once the field is established~\cite{newman2009first,van2015interdisciplinary}.

\subsection*{Recognition of physics subfields}
Do impact differences affect the way in which the overall scientific community perceives the different subfields of physics? As a rough proxy of this recognition we take the Nobel Prizes awarded from 1985 to the present, highlighting each awarded subfield (Fig.~\ref{fig:impact}j, SI Section S10). Although the Nobel Prize often recognises research undertaken much before the selection year, the timing of Nobel prize selections could affect the way in which the relative importance of different physics communities are perceived by the committee. As a comparison between Fig.~\ref{fig:transition}a and Fig.~\ref{fig:impact}j shows, Nobel Prizes are not related to the number of physicists flocking into specific physics communities, nor do they show significant temporal clusters. However, the general distribution of awarded subfields reveals interesting tendencies: a large fraction of Nobel Prizes have been awarded to the ``curricular'' topics, like {\sf CondMat}, the subfield with the largest number of active researchers, and {\sf HEP}. Surprisingly, {\sf Astro}, despite the relatively moderate size of its community, comes in third, with five Nobel Prizes. This success might be linked to the perception of astrophysics as a field that studies the universe on a grand scale, as well as to its strong ties to HEP, a regular recipient of Nobels. Other well established areas with a long history, such as {\sf AMO} and {\sf Classical} have also been recognised. In contrast, since 1985 {\sf Plasma} and {\sf Interdisc} have not been awarded a Nobel Prize. The omission of {\sf Interdisc} likely comes from the charter of the Nobel Prize to award clear-cut categories (e.g. physics, chemistry, medicine/physiology) rooted in 19th century discriminating against interdisciplinary discoveries~\cite{szell2017Interdisciplinarity,bromham2016funding}.

\subsection*{Conclusions}
As one of the oldest scientific disciplines, physics plays a fundamental role in the development of science. As the aperture of physics widens, the focus of individual physicists narrows, leading progressively to the formation of specialised communities and subfields. Here we offered an intellectual census of these subfields, exploring how physicists migrate between them, how they specialise and collaborate to create impactful research.

We observed that subfields rarely live in isolation but rather tend to overlap, with individual scientists working in multiple subfields and transitioning between fields during their career. Mapping these overlaps reveals a highly non-trivial research space, displaying deep intellectual links between some subfields and large gaps between others.  

Physicists who are confronted with heated arguments on the allocation of resources to different subfields and departments, often use metrics of productivity or impact to seek priority. However, our research suggests that such arguments should be taken with scepticism. Indeed, there are considerable field-specific differences in the patterns of productivity and impact.
Publication rates have exploded in recent years in {\sf HEP}, {\sf Nuclear} and {\sf Astro}, whereas fractional productivity is declining. In some subfields, such as {\sf HEP}, researchers co-authors an exceptionally large number of papers, partly rooted in their unique culture of collaboration. By contrast, interdisciplinary physicists produce papers at a much lower rate but their papers tend to garner a disproportionally higher impact, once we factor in the relative size of the subfield. 
Understanding these field differences within physics represents the first step towards a deeper understanding of our discipline. As tomorrow's physicists working on different topics compete for the same position and resources, these insights may prove pertinent for the sustainable vitality of physics as a discipline.

Our study is based on Web of Science data, lacking the literature that has been exclusively published in preprint servers like arXiv\cite{arXivdataset}, leading to unavoidable (but small) differences in subfield representation due to diverse publication cultures in different communities. For example, the proportion of {\sf HEP} and {\sf Astro} papers in arXiv is higher compared to our dataset and WoS, reflecting the common practice of these communities to communicate findings in preprints rather than journal papers. However, there is a high overlap in the coverage of the physics literature between different databases \cite{martin2018coverage}
and a high correlation of the representation of physics subfields (SI Section S3), indicating that our findings should agree if repeated on a different database. 

In this study we focused on careers of physicists within physics. However, these days, many scientists with a background in the physical sciences contribute to fields outside of physics, from biology to finance, both in academia and the private sectors~\cite{farmer1999physicists}. For this reason, the investigation of the connection between physics and other scientific disciplines, and the career transitions away from physics, remains as fruitful future work. Indeed, such an investigation, possibly aided with data sources that go beyond scientific publications, could shed light on the role of physics and its subfields in the entire ecosystem of science and beyond.  

\subsection*{Acknowledgments}
This work was supported by the John Templeton Foundation Grant \#61066  (A.-L.B., F.B., R.S., and M.S.), the ITI project ``Just Data'' funded by Central European University (F.M. and R.S.), the National Science Foundation grant SBE 1829344 (D.W.), and the Air Force Office of Scientific Research grants FA9550-15-1-0077 (A.-L.B., R.S., and M.S.), FA9550-15-1-0364 (A.-L.B. and R.S.), FA9550-15-1-0162 (D.W.) and FA9550-17-1-0089 (D.W.).

\clearpage

\begin{figure}[ht]
\centering
\includegraphics[width=0.8\textwidth]{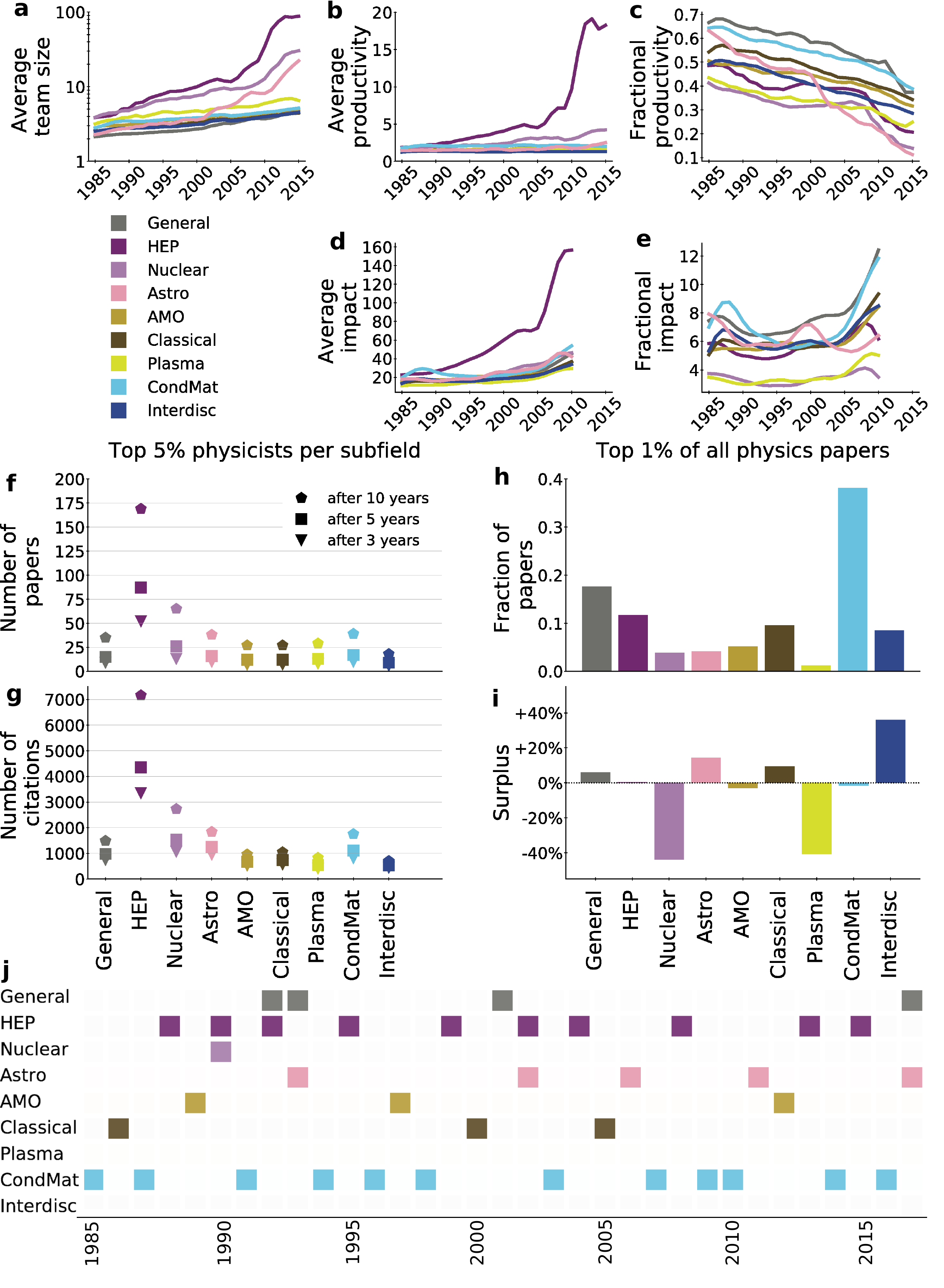}
\end{figure}
\clearpage
\captionof{figure}{\textbf{Productivity and impact across physics communities.} \textbf{a}, Average team size, defined as average number of authors per paper, over time. Team sizes grow in all fields, especially in {\sf HEP}, {\sf Nuclear}, and {\sf Astro} due to large-scale experimental projects. \textbf{b}, Average productivity, defined as number of papers per author, over time. Productivity grows for {\sf HEP}, {\sf Nuclear}, and {\sf Astro} but stays roughly constant for other subfields. \textbf{c}, Fractional productivity, i.e.~number of papers divided by team size, over time. For all subfields productivity grows less than team size, therefore fractional productivity decreases. \textbf{d}, Average impact, defined as number of citations per author within a 5 years window. Impact increases in all fields, but only {\sf HEP} shows an exceptional growth. \textbf{e}, Fractional impact, i.e. number of paper citations divided by team size, over time. Most subfields show a roughly constant trend until 2005. \textbf{f}, Number of papers of the top $5\%$ physicists for productivity. Due to different collaboration standards, {\sf HEP} physicists coauthor more papers than other subfields. {\sf Interdisc} physicists produce an especially low number of papers. \textbf{g}, Number of citations of the top $5\%$ physicists for productivity. {\sf HEP} physicists receive more citations because of their high productivity. \textbf{h}, Fraction of top 1\% cited papers per subfield and \textbf{i}, subfield surplus with respect to the number expected given the subfield size. {\sf Interdisc} generates the highest number of high impact papers compared to its size. \textbf{j}, Nobel Prizes in physics per year across subfields. {\sf Plasma} and {\sf Interdisc} have not received an award.} 
\label{fig:impact}

\clearpage

\bibliography{success_natrevphys.bib}

\end{document}


\maketitle

\thispagestyle{empty}

\section{Defining physics publications in non-physics journals}
\label{sectionSI:physics}
We identify physics publications in journals which are not explicitly labelled as physics journals by means of a method first used in Refs.~\cite{Sinatra2015, DevilleThesis}. Such method allows to reconstruct a community in a network when only a small fraction of nodes are explicitly labelled as belonging to the community.
In our case, the hypothesis is that physics papers can be found not only in conventional physics journals (core physics papers) but also in other venues (interdisciplinary physics papers).
It is possible to identify such interdisciplinary papers if they have a significant number of references or citations in conventional physics venues. In Ref.~\cite{Sinatra2015} the label propagation algorithm was first applied to an old version of the Web of Science (WoS), encoding information about scientific publications until 2012 and based on an old database structure. Here we reapply the method on an updated version of WoS purchased from Clarivate Analytics, encoding information about publications until 2017, and using a new database structure, with a different identification system for papers among other things. We obtain a new physics dataset of papers, which we want to further characterise by identifying the physics subfields they belong to. For this reason, papers in the dataset except those of the American Physical Society (APS) journals, are then considered to be assigned a given subfield and be part of our physics communities analysis. The label propagation method at the subfield level is a modified implementation of the algorithm presented in this Section, and it is illustrated in detail in Section \ref{sectionSI:subfields}.

The label propagation method to construct the physics dataset works in the following way. Let us consider a directed network with $N$ nodes, for instance the citation network described by the WoS dataset, where nodes are scientific publications, and a direct link between publication $i$ and publication $j$ exists if paper $i$ cites paper $j$. Each node $i$ has an in-degree $k_{IN}$ (number of citations) and an out-degree $k_{OUT}$ (number of references). Nodes with $k_{IN}=0$ and $k_{OUT}=0$ are publications without references and citations and are isolated nodes in the network. Additionally, in our case each node $i$ is characterised by a variable $t_i$ corresponding to the time of publication of the article. The method is based on an iterative process where at each step $s$ the $N$ nodes are assigned to three sets: the core set $C_s$, the tangent set $T_s$ and the external set $E_s$.
The core set $C_s$ includes the nodes that are considered to be part of the target community at a given time step $s$ by the algorithm. In our case, at the step $s=0$, $C_0$ includes all articles published in physics journals. The purpose of this initial core set is to act as a seed to detect other nodes that are part of the community, even if initially they are not classified as such, and that will be iteratively included in $C_s$ at subsequent steps $s$ > 0. The second set is the tangent set $T_s$, and contains all the nodes outside the core set $C_s$ that have at least one (ingoing or outgoing) connection to a node within $C_s$. The third set is the external set $E_s$, and corresponds to all nodes outside the core set $C_s$ that share no connection with nodes within $C_s$, and therefore have no chance to be included into the core at the subsequent step $s+1$. By definition we have $C_s \cup T_s \cup E_s = N$ and $C_s \cap T_s \cap E_s = \emptyset$.

The basic idea of the method is to iteratively extend the target community $C_s$ into $C_{s+1}$ by adding candidate nodes from $T_s$ that are statistically expected to be part of the community based on their connections. In our case this corresponds to identifying as physics all scientific papers which are not published in physics journals, but whose patterns of references and citations are indistinguishable from those published in the traditional physics venues. The purpose of the tangent set $T_s$ is to contain all candidate nodes, i.e. nodes that might subsequently
be added to the target community $C_s$ at step $s$ after inspection of their incoming and outgoing links. To do so, at each step $s$ and for each node $i$ we compute two variables: $r_{i,s}^{IN}$ and $r_{i,s}^{OUT}$. These variables quantify the expectation of a particular node to be part of the target community $C_s$ based on its incoming citations and outgoing references.

Let us focus first on incoming citations, evaluated through $r_{i,s}^{IN}$, where 
\begin{equation}
r_{i,s}^{IN} =\frac{k_{i,s}^{IN, \bigodot}} { \hat k_{i,s}^{IN, \bigodot}   }.
\end{equation}
Here $k_{i,s}^{IN, \bigodot}$ corresponds to the number of incoming links (citations) to node $i$ originating from nodes in the core $C_s$. $\hat k_{i,s}^{IN, \bigodot}$, instead, accounts for the expected number of incoming links from the core in a null model where the real number of incoming and outgoing links of each node (citations and references of each paper) in the network is fixed. This last constraint corresponds to consider the directed configuration model ensemble of the original citation network, meaning that we can write
\begin{equation}
\hat k_{i,s}^{IN, \bigodot} =  k_i^{IN} \frac{\sum_{j \in  C_s }  k_j^{OUT}      }    { \sum_{j \in N}  k_j^{OUT}    }
\end{equation}
where $k_i^{IN}$ denotes the total number of incoming links to node $i$, and the remaining term corresponds to the probability for a link to originate from $C_s$. 
As an article $i$ can receive a citation from another paper $j$ only if the latter is more recent, i.e. $t_j>t_i$, we eventually set  
\begin{equation}
\hat k_{i,s}^{IN, \bigodot} =  k_i^{IN} \frac{\sum_{j \in  C_s | t_j>t_i}  k_j^{OUT}      }    { \sum_{j \in N | t_j>t_i}  k_j^{OUT}    }.
\end{equation}

Similarly, the share of outgoing references are evaluated through $r_{i,s}^{OUT}$, where 
\begin{equation}
r_{i,s}^{OUT} =\frac{k_{i,s}^{OUT, \bigodot}} { \hat k_{i,s}^{OUT, \bigodot}   },
\end{equation}
and
\begin{equation}
\hat k_{i,s}^{OUT, \bigodot} =  k_i^{OUT} \frac{\sum_{j \in  C_s | t_j<t_i}  k_j^{IN}      }    { \sum_{j \in N | t_j<t_i}  k_j^{IN}    }.
\end{equation}

A value $r_{i,s}^{IN} > 1$  ($r_{i,s}^{OUT}$>1) corresponds to a node that is more likely to reference (be cited from) nodes from the core than what would be expected at random. At each step $s$ of the process, we use the variables $r^{IN}_{i,s}$ and $r^{OUT}_{i,s}$ associated to nodes in $T_s$ to produce the updated core set
$C_{s+1}$. First we add all nodes in $C_{s}$ to $C_{s+1}$. Then, for each node $i \in T_s$, we add $i$ to $C_{s+1}$ if we have  
\begin{equation} 
r_{i,s}^{IN} > \tau^{IN}
\label{eq: tauin}
\end{equation} 
or
\begin{equation}
r_{i,s}^{OUT} > \tau^{OUT}.
\label{eq: tauout}
\end{equation}
The thresholds $\tau^{IN}$ and $\tau^{OUT}$ are fixed based on a parameter $p$ such that the thresholds $\tau^{IN}$ and $\tau^{OUT}$ correspond respectively to the $p-th$ percentile
of the distribution of $r_{i,0}^{IN}$ and $r_{i,0}^{OUT}$ values for nodes within the initial core set $C_0$. 
Once nodes $i \in T_s$ satisfying the conditions of Eq.\ref{eq: tauin} or  Eq.\ref{eq: tauout} are
added to the core set $C_{s+1}$, both sets $T_s$ and $E_s$ can be updated to $T_{s+1}$
and $E_{s+1}$ from $C_{s+1}$. The process stops when $C_s$ has converged, i.e. when no nodes
from $T_s$ can be added to the core set $C_s$. Note that while the thresholds
$\tau^{IN}$ and $\tau^{OUT}$ remain constant during the whole process, the values $r_{i,s}^{IN}$ and $r_{i,s}^{OUT}$ associated to each node $i$ will change at each iteration, given the fact that new nodes will incorporate the set $C_s$ at each iteration step s. As shown in Ref.~\cite{DevilleThesis}, in the case of physics publication in the WoS dataset the algorithm was run iteratively for $10$ steps, showing fast convergence.

\begin{table}[!ht]
\small

\centering
\begin{center}

 \begin{tabular}{||c | c | c ||} 
 \hline
 \textbf{Rank} & \textbf{Journal \footnotesize{(number of papers)}} & \textbf{Journal \footnotesize{(percentage of papers)} }  \\ 
 \hline\hline
  1 & Rev. Sci. Instrum. (41,006)   & J. Space Weather Space Clim. (98.3\%)\\ 
 \hline
   2 & Thin Solid Films (38,316) & Quantum Inform. Comput. (97.7\%)\\ 
 \hline
   3 & Surf. Sci. (34,461) & J. Hyberbolic Differ. Equ. (93.2\%)\\ 
 \hline
   4 & Nature (33,073) & Adv. Quantum Chem. (92.8\%)\\ 
 \hline
   5 & J. Alloy. Compd. (31,319) & 2D Mater. (92.7\%) \\ 
 \hline
   6 & J. Phys. Chem. (29,920) & Thin Solid Films (92.2\%)\\ 
 \hline
    7 & J. Am. Chem. Soc. (27,192) & J. Laser Micro Nanoeng. (92.2\%)\\ 
 \hline
    8 & Macromolecules (26,377) & Surf. Sci. Rep. (91.9\%)\\ 
 \hline
    9 & Electron. Lett. (25,593) &  IEEE Trans. Nanotechnol. (91.4\%)\\ 
 \hline
    10 & J. Electrochem. Soc. (24,806) & Symmetry Integr. Geom. (90.1\%)\\ 
 \hline
\end{tabular}

\hspace{0.3in}

\end{center}

\caption{\textbf{Non-physics journals with most physics publications and highest percentage of physics publications identified by means of label propagation.}}
\label{tab:nonphysicsjournals}
\end{table}

The parameter $p$ can be considered as a tolerance parameter in the sense that it defines the
minimal attraction needed for a node to be incorporated in the growing
core. As described in Refs.~\cite{Sinatra2015, DevilleThesis}, in our case it is possible to set the value of $p$ by validating the algorithm on all publications of two interdisciplinary journals for which a subset is labelled explicitly as physics, namely \textit{Science} (1995-2013) and \textit{PNAS} (1915-2013). The best trade-off between true positive ($92.3\%$) and true negative rates ($99.6\%$) was found for $p=10$.  By running the algorithm on the new version of the WoS dataset comprised of ${\sim}$54 million papers, with an initial core of ${\sim}$3.2 million articles published in 294 physics journals, the list of journals being extracted by combining information from Wikipedia, Scopus and Scimago, we identified an additional number of ${\sim}$4.5 million physics publications in non-physics journals. In Table~\ref{tab:nonphysicsjournals} we report the ten non-physics journals with the highest number of physics publications (number of papers in brackets), and the ten non-physics journals with the highest share of physics publications (percentages in brackets). We note the presence of interdiscisciplinary journals, such as \textit{Nature}, and several materials and chemistry journals.


\section{Identifying physics subfields from PACS codes}
\label{sectionSI:pacs}

Despite the WoS dataset provides a thorough classification of core physics publications into different subfields (see Section \ref{sectionSI:subfields}), such classification is not detailed enough to our scope and, most importantly, it fails to associate a subfield to publications not in physics journals. For such a reason, in our work we associated publications to different subfields according to the Physics and Astronomy Classification Scheme (PACS) by the American Physical Society~\cite{PacsAPS}, a hierarchical classification used for papers in APS journals between 1977 and 2015. The classification uses four digits and an extra identifier. The 1-digit identifies 10 different physics subfields, namely: {\sf General} (0), The Physics of Elementary Particles and Fields (shortened as {\sf HEP}, 1), {\sf Nuclear} Physics (2), Atomic and Molecular Physics ({\sf AMO}, 3), Electromagnetism, Optics, Acoustics, Heat Transfer, Classical Mechanics, and Fluid Dynamics ({\sf Classical}, 4),  Physics of Gases, Plasmas, and Electric Discharges ({\sf Plasma}, 5), Condensed Matter: Structural, Mechanical and Thermal Properties (6), Condensed Matter: Electronic Structure, Electrical, Magnetic, and Optical Properties (7),  Interdisciplinary Physics and Related Areas of Science and Technology ({\sf Interdisc}, 8), Geophysics, Astronomy, and Astrophysics ({\sf Astro}, 9). We merged PACS 6 and 7 into a unique category named {\sf CondMat}, in order to match other common physics classifications, such as that found for the arXiv (see Section~\ref{sectionSI:subfields}). We stress that the term interdisciplinary physics, assigned in Ref.~\cite{Sinatra2015} to describe physics publications in non-physics journals, is not linked to the PACS 8 of the APS scheme. In the following, as well as in the main text, the term Interdisciplinary physics is reserved to identify publications and authors working in this precise subfield of physics, differently from Ref.~\cite{Sinatra2015}. PACS can be found in the APS dataset, available from the APS upon request~\cite{APSdataset}, encoding information about all publications appeared in the journals of the American Physical Society until 2015. Although PACS appeared in 1977, only a small fraction of the papers were assigned one until they were enforced in 1985. For this reason, we focused our analysis on the years 1985-2015, for which our dataset has 435,722 papers with at least one PACS. 5,616 more papers have assigned a PACS but were published before 1985. More in detail, between 1985-2015 we have 265,549 papers with exactly one 1-digit PACS, 138,176 with two PACS, 29,806 with three PACS, 2,160 with  PACS and 31 with five PACS. 

\begin{figure}[ht]
\centering
\includegraphics[width=0.8\textwidth]{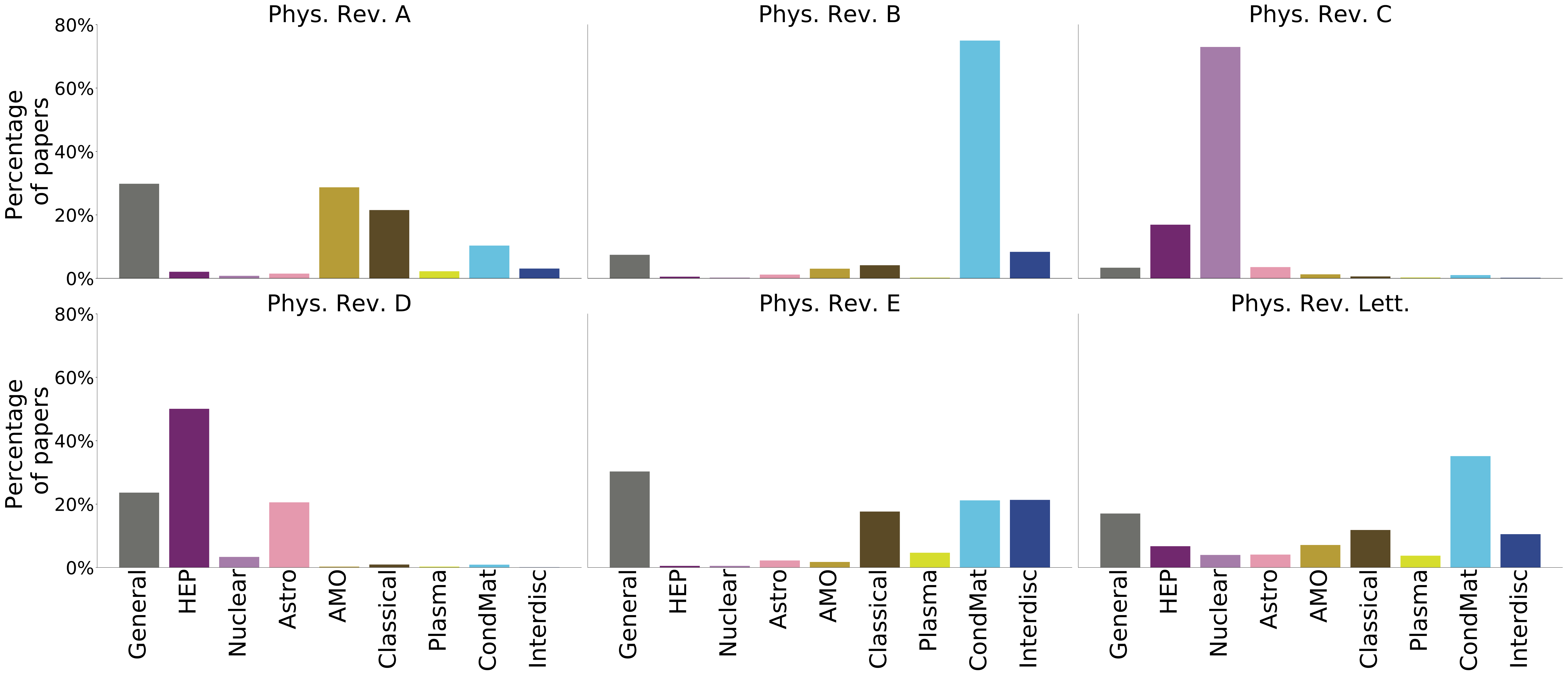}
\caption{\textbf{Subfield distribution for papers published in APS journals.} Different APS journals show different publication patterns across subfields.  \textit{Physical Review B} covers predominantly {\sf CondMat}, and \textit{Physical Review C} is similarly focused on {\sf Nuclear}. In contrast, \textit{Physical Review A}, \textit{Physical Review D} and \textit{Physical Review E} do not cover a single, predominant subfield. \textit{Physical Review Letters} is the most balanced journals of the APS publishing across all subfields.}
\label{fig:paperpacs_APS_journals}
\end{figure}

In Fig.~\ref{fig:paperpacs_APS_journals} we report the distribution of the 9 physics subfields for six well-established journals published by the APS, namely the general purpose \textit{Physical Review Letters} and the specialised venues \textit{Physical Review A - E}. \textit{Physical Review B} (covering condensed matter and materials physics) and \textit{Physical Review C} (covering nuclear physics) indeed predominantly publish papers belonging to a single subfield, respectively Condensed Matter and Nuclear Physics. Conversely \textit{Physical Review A} (covering atomic, molecular, and optical physics and quantum information), \textit{Physical Review D} (covering particles, fields, gravitation, and cosmology) and \textit{Physical Review E} (covering statistical, nonlinear, biological, and soft matter physics) publish across a greater mixture of subfields. As expected, \textit{Physical Review Letters}, the APS flagship journal, publishes across all different domains, even though with different frequency. 

Similarly to the identification of physics papers in non-physics venues, we use the papers published in the APS journals as the initial seed to assign subfields to other physics publications by means of label propagation (see Section~\ref{sectionSI:subfields} for details.). In such a way, we obtain a data-driven subfield classification of physics papers in the WoS dataset.

In Fig.\ref{fig:APS_WoS_comparison}a we report the proportions of APS papers belonging to a given subfield, and compare it to that of our newly created dataset.
In Fig.\ref{fig:APS_WoS_comparison}b we report the distribution of the number of subfield per paper in the APS between 1985 and 2015, as well as the fraction of number of papers per subfield over the years (Fig.\ref{fig:APS_WoS_comparison}c).

\section{Assigning Physics subfields to Web of Science publications}
\label{sectionSI:subfields}

We propagate physics subfields to physics publications in the WoS dataset based on relevant patterns of references and citations to the specific subfield(s), adapting the method described in the first section of this SI. For each subfield we have a different initial core set $C_0^{\alpha}$, corresponding to all publications in the APS publications between 1985 and 2015 associated to a given subfield $\alpha$. 
First, we matched the papers of the APS dataset into the Web of Science dataset, either via exact doi matching, or, for when the doi is not available, by using the Levenshtein distance to compute title similarity. In this second case the match was accepted if there was at least 90\% string similarity between the titles of two papers in the datasets, and the second best match had a string similarity at least 5 times worse. In this way we were able to match 90\% of all the papers manually assigned to a subfield between 1985 and 2015.

At difference with the original implementation, where it was possible to set the thresholds $\tau^{IN}$ and $\tau^{OUT}$ by evaluating the performance of the algorithm on the 'groundtruth' of physics papers published in interdisciplinary journals such as \textit{Science} and \textit{PNAS}, such type of validation is not possible at the subfield level. For such a reason, for label propagation at the subfield level we slightly modified the original implementation. We observe that the algorithm may propagate subfields both to papers within and out of the original APS core, which is made of papers that already have a PACS code. For such a reason, for each subfield $\alpha$ we selected the threshold $\tau^{\alpha}$ so that after $10$ iterations the number of papers of each subfield cannot grow more than 10\% within the original APS dataset. For simplicity, we chose $\tau^{IN, \alpha} = \tau^{OUT, \alpha}$. Afterwards, we performed label propagation for each subfield $\alpha$ independently. We obtained a total of 1,137,670 papers in WoS published between 1985 and 2015 and classified within one of the subfields of Physics. We note that also some papers outside the considered time-span were assigned a subfield, but we focused our analysis on the period $1985-2015$ to be consistent with the years when PACS were systematically used in publications by the APS. As already mentioned, PACS corresponding to the two categories associated to Condensed Matter were merged into the same subfield.

\begin{figure}[ht]
\centering
\includegraphics[width=0.8\textwidth]{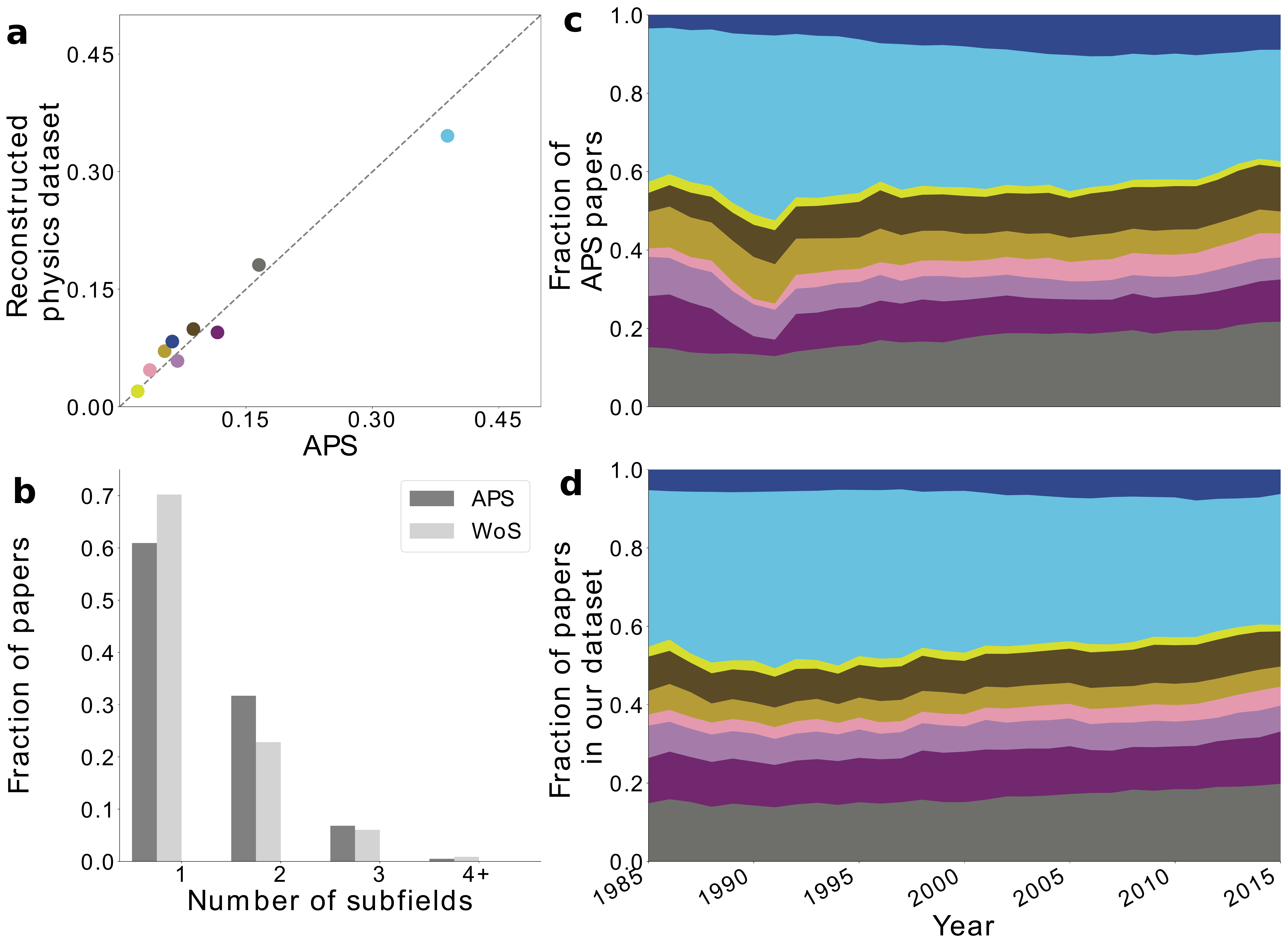}
\caption{\textbf{Comparison between the APS dataset and the reconstructed physics dataset.} \textbf{a} Scatterplot of the fraction of subfields appearing in papers of the APS dataset and in the reconstructed physics dataset. \textbf{b} Distribution of number of subfields per paper in the two datasets. \textbf{c}, \textbf{d} Temporal evolution of the fraction of subfields between 1985 and 2015 for the two datasets.}
\label{fig:APS_WoS_comparison}
\end{figure}

It is interesting to compare the classification of papers obtained through label propagation with that of the original APS dataset. Figure~\ref{fig:APS_WoS_comparison}a compares the fraction of subfields in the original and the propagated datasets. The two datasets have a similar subfield distribution with a cosine similarity of 0.99. Differences in the two datasets are likely to indicate an under- or over- representation of some areas of physics in the \textit{Physical Review} series compared to the overall physics world. In Fig.~\ref{fig:APS_WoS_comparison}b we report the distribution of the number of subfields per paper in the two datasets. Papers in the reconstructed physics dataset tend to be slightly more specialised ($70\%$ of the papers are assigned to a single subfield) than those in the APS dataset ($61\%$). However, overall the two distributions are quite similar. Finally, in Figs.~\ref{fig:APS_WoS_comparison}c,d we show the evolution of the fraction of papers of different subfields in the APS dataset and in our reconstructed dataset from 1985 to 2015. It is evident how the two datasets have very similar temporal patterns during the period under investigation. 

\begin{figure}[ht]
\centering
\includegraphics[width=0.8\textwidth]{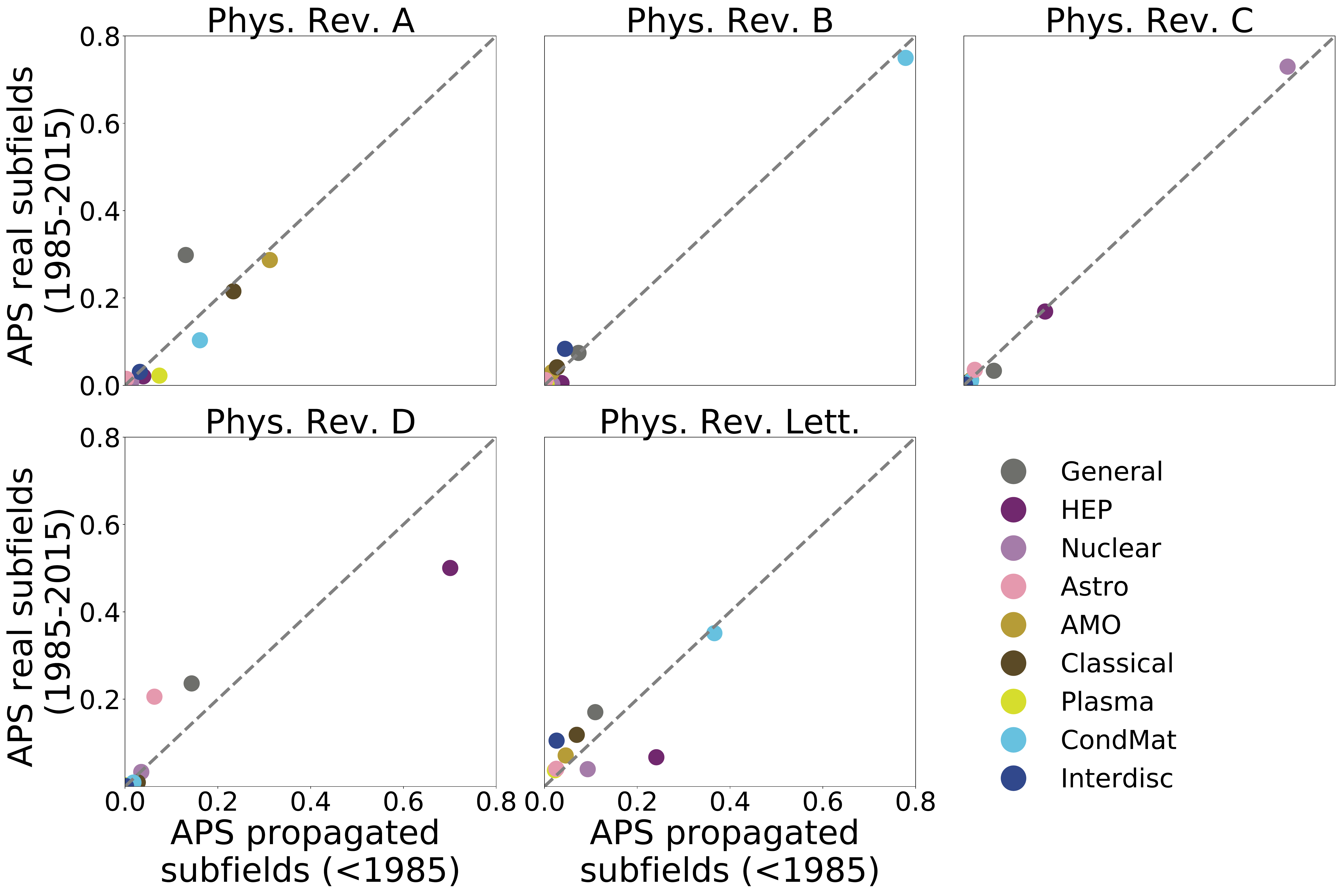}
\caption{\textbf{Testing propagated subfields in APS journals before 1985.} Scatterplot between the subfield distribution of the papers published in the APS journals after 1985, and the propagated subfield distribution for papers published before 1985 in the same journals. The cosine similarities between the distribution of papers before and after 1985 are \textit{(i)} 0.91 for \textit{Physical Review A}, \textit{(ii)} 0.99 for \textit{Physical Review B}, \textit{(iii)} 0.99 for \textit{Physical Review C}, \textit{(iv)} 0.93 for \textit{Physical Review D} and \textit{(v)} 0.88 for \textit{Physical Review Letters}.}
\label{fig:APS_WoS_scatter}
\end{figure}

{\bf Validation:} To test the robustness of our findings, we validated our data-driven classification of papers across subfields. As already mentioned, PACS codes were systematically introduced in publications in the APS journals 1985. As our method classifies papers into subfields according to patterns of references and citations only, our algorithm naturally assigns subfields also to publications in the APS journals before 1985, provided that they are significantly connected to the corresponding core papers for the subfield(s). Five of the previously six analysed APS journals (with the exception of \textit{Physical Review E}) were born before 1985. In Fig.~\ref{fig:APS_WoS_scatter} we test the robustness of the subfield distributions in the journals  as a way to assess the effectiveness of our data-driven method to classify physics papers across subfields by comparing the distribution of the subfield manually assigned between 1985 and 2015 in \textit{Physical Review. A, B, C, D}, and \textit{Physical Review Letters}, with that obtained by means of label propagation for papers published before 1985 in the same journals. The two distributions are highly correlated for all journals, with cosine similarities ranging from 0.88 to 0.99 . 

\begin{figure}[ht]
\centering
\includegraphics[width=0.8\textwidth]{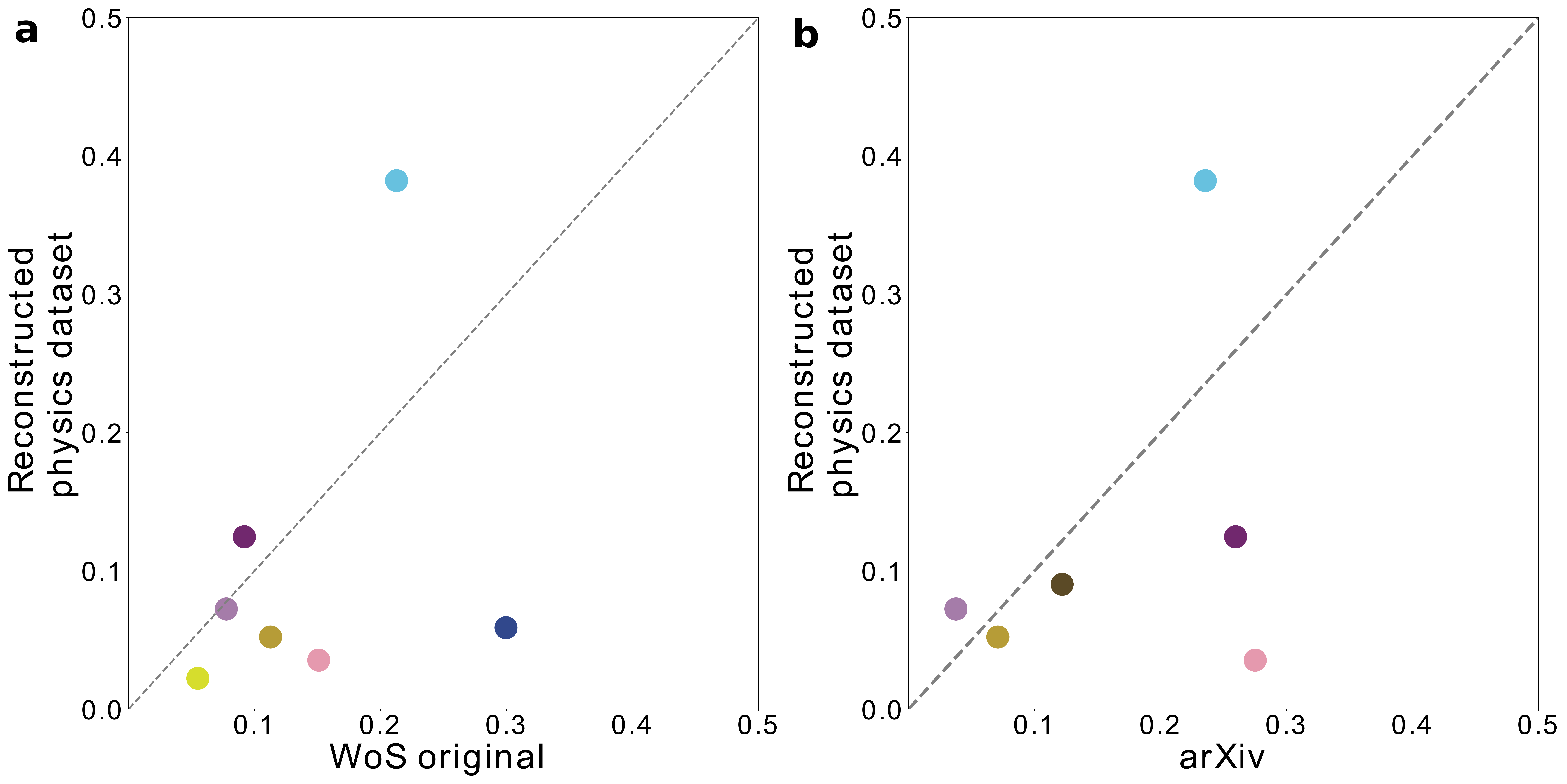}
\caption{\textbf{Comparison between the distribution of subfields in our reconstructed physics dataset with the WoS and the arXiv physics categories.} Correlation between distributions is high, with values of cosine similarity respectively equal to \textbf{a} 0.86, and \textbf{b} 0.74.}
\label{fig:val_papers}
\end{figure}

We also tested the robustness of our subfield categorisation by comparing it to additional sources providing alternative physics classifications, namely the physics classification provided by
 \textit{(i)} the WoS dataset (for core physics papers only), \text{(ii)} the arXiv repository, that collects electronic preprints of papers related to physics topics. The cosine similarity between the fraction of papers in our dataset and in the two alternative datasets is quite high, respectively \textit{(i)} 0.86 for WoS, \textit{(ii)} 0.74 the arXiv. The scatterplots between our reconstructed physics dataset and the other databases are shown respectively in Figs.~\ref{fig:val_papers}a,b. Achieving a perfect mapping between the scheme of arXiv and WoS into the APS scheme is not possible. As an example, the \textit{nonlin} category in the arXiv dataset, that we eventually mapped into the {\sf General} physics subfield, actually contains papers of at least an additional subfield, i.e. {\sf Interdisc}. For the same reason some of the subfields obtained from the PACS scheme do not have a direct counterpart in the other two datasets. We report the full mappings in Table~\ref{tab:comparison_papers}. 

\begin{table}

\begin{center}
\small

 \begin{tabular}{||c | c | c||} 
 \hline
 \textbf{WoS category} & \textbf{Subfield} & \textbf{arXiv category} \\ 
 \hline\hline
 / & {\sf General} & / \\ 
 \hline
  Fields  & {\sf HEP} & hep-ex, hep-lat, hep-ph, hep-th, math-ph   \\
 \hline
 Nuclear Physics &  {\sf Nuclear} & nucl-ex, nucl-th \\
   \hline
 Astrophysics &  {\sf Astro} & astro-ph, gr-qc \\
   \hline
  Atomic, Molecular \& Chemical Physics & {\sf AMO} &  quant-ph \\
   \hline
  / & {\sf Classical} & physics, nlin \\
   \hline
  Fluids \& Plasmas Physics & {\sf Plasma} & / \\
   \hline
  Condensed Matter Physics & {\sf CondMat} & cond-mat \\
  \hline
  Multidisciplinary Physics & {\sf Interdisc} & / \\
 \hline
\end{tabular}


\end{center}
\caption{\textbf{Mapping of physics categories from arXiv categories and WoS physics categories into physics subfields.}}
\label{tab:comparison_papers}
\end{table}

Another factor that may affect the matching is the presence of specific biases for each of these datasets, which are captured by comparing it with our new data-driven reconstructed physics dataset. For instance, the arXiv, first created as a repository for people working on High Energy Physics, shows a disproportionally high number of {\sf HEP} and {\sf Astro} publications. This comes as no surprise since the initial scope of the arXiv was to diffuse scientific results in {\sf HEP}, and the repository has been largely used by such community. 

In Table~\ref{tab:PACSjournals}, we report the five non-APS journals with most papers assigned to each subfield by means of label propagation (number of papers in brackets).

\begin{table}[!ht]
\footnotesize

\centering
\begin{center}
 \begin{tabular}{||c | c | c | c ||} 
 \hline
 \textbf{Rank} & \textbf{General} & \textbf{HEP} & \textbf{Nuclear} \\ 
 \hline\hline
  1 & Phys. Lett. A (9,027) & Nucl. Phys. B (14,524) & Nucl. Phys. A (16,680)  \\ 
 \hline
   2 & J. Phys. A-Math. Gen. (6,029) & J. High Energy Phys. (10,860) & Phys. Lett. B (9,017) \\ 
 \hline
   3 & Physica A  (4,863) & Nucl. Phys. A (7,109) & J. Phys. G-Nucl. Part. Phys. (4,916)  \\ 
 \hline
   4 & Class. Quantum Gravity (4,299) & Prog. Theor. Phys. (5,280) & Nucl. Instrum. Methods Phys. A (4,238)  \\ 
 \hline
   5 & J. Math. Phys. (3,366) & Eur. Phys. J. C (4,462) & Eur. Phys. J. A (2,848)  \\ 
 \hline
\end{tabular}
\end{center}

\vspace{0.3in}

\begin{center}
 \begin{tabular}{||c | c | c | c||} 
 \hline
 \textbf{Rank} & \textbf{Astro} & \textbf{AMO} & \textbf{Classical} \\ 
 \hline\hline
  1 & Phys. Lett. B (3,559) & J. Phys. B (7,005) & Opt. Commun. (4,179)  \\ 
 \hline
   2 & J. Cosmol. Astropart. Phys. (2,370) & J. Chem. Phys. (2,314) & Phys. Lett. A (3,955) \\ 
 \hline
   3 & Astrophys. J. (2,330) & Nucl. Instrum. Methods Phys. B (1,483) & Opt. Lett. (2,492) \\ 
 \hline
   4 & Class. Quantum Gravity (2,099) & Phys. Lett. A (1,460) & J. Opt. Soc. Am. B (2,319) \\ 
 \hline
   5 & Nucl. Phys. B (1,219) & Phys. Scr (1,089) & J. Appl. Phys. (2,123) \\ 
 \hline
\end{tabular}
\end{center}

\vspace{0.3in}

\begin{center}
 \begin{tabular}{||c | c | c | c||} 
 \hline
 \textbf{Rank} & \textbf{Plasma} & \textbf{CondMat} & \textbf{Interdisc} \\ 
 \hline\hline
  1 & Phys. Plasmas (2,749) & J. Appl. Phys. (23,364) & Physica A (1,977)  \\ 
 \hline
   2 & Phys. Fluids (1,088) & Appl. Phys. Lett. (20,196) & J. Phys. A-Math. Gen. (1,923) \\ 
 \hline
   3 & Rev. Sci. Instrum (908) & Physica C (19686) & J. Chem. Phys. (1,636) \\ 
 \hline
   4 & Nucl. Instrum. Methods Phys. A (873) & Solid State Commun. (16,274) & Phys. Lett. A (1,419) \\ 
 \hline
   5 & Plasma Phys. Control. Fusion (823) & Physica B (15,247) & J. Appl. Phys. (1,297) \\ 
 \hline
\end{tabular}

\end{center}
\caption{\textbf{Non-APS journals with most publications with propagated subfields.}}
\label{tab:PACSjournals}
\end{table}

We note that the Astrophysics literature seems to be relatively disconnected to its APS core, compared to results for the other subfields. As an example, we focus on a well established specialised journal in the area, the \textit{Astrophysical Journal}, for which WoS indexes 98,482 papers, only 2,330 of which are labeled. This is because, out of the 3,724,542 outgoing references from papers published in the \textit{Astrophysical Journal}, only $0.6\%$ are directed towards the {\sf Astro} core. Similarly, out of the 4,896,146 incoming citations towards papers published in the \textit{Astrophysical Journal}, only $1.4\%$ come from the {\sf Astro} core.  As a reference, we compare these numbers with those of \textit{Solid State Communications}, a specialised journal in the area of Condensed Matter, for which our method assign a subfield to 16,274 out of 35,781 papers. In such case, of the 489,625 references and 635,466 citations of the journal, $5.3\%$ and $4.8\%$ link to the {\sf CondMat} core. These numbers are roughly fives times higher than those for the \textit{Astrophysical Journal}. As a consequence of this disconnection, it is possible that our method it is underestimating the number of (possibly specialised) scientists working in Astrophysics. For both journals the fraction of citations (references) coming from (going towards) the cores associated to the other subfields is negligible.

\begin{figure}[ht]
\centering
\includegraphics[width=0.8\textwidth]{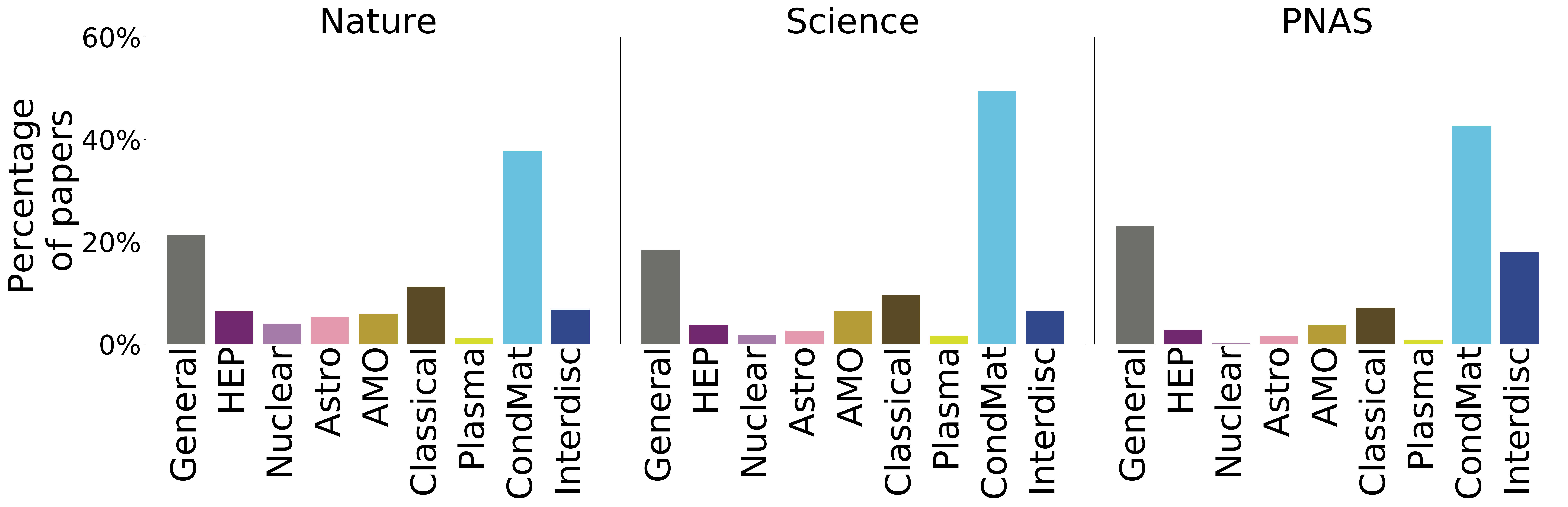}
\caption{\textbf{Shares of subfields for publications in \textit{Nature}, \textit{Science} and \textit{PNAS}.} All three interdisciplinary journals publish across all subfields of physics.}
\label{fig:interjournals}
\end{figure}

At last, in Fig.~\ref{fig:interjournals} we report the publication profile across subfields for three leading interdisciplinary journals. Unsurprisingly, most subfields are represented in all three venues. We note that the proportions of the different subfields is similar to that of the publication of the APS flagship journal, \textit{Physical Review Letters}.

\clearpage

\section{Assigning physicists to subfield(s)}
\label{sectionSI:authors}

While papers are directly associated to subfields through label propagation, we still need to assign physicists to their correct research area. Some physicists, in particular those extremely productive, are likely to appear over a whole career as the authors of papers belonging to multiple subfields, though some of these might not be significant. As a consequence, when assigning the authors to the different subfields, we applied a statistical filter in order to assign only the subfield(s) on which their engagement is significant. In particular, we consider a physicist as significantly working in a subfield only if her share of publications in it, compared to her production across all subfields, is greater than that of the average scientist. Let us consider the bipartite weighted network $W = \{ w_{i \alpha} \}$, where $w_{i \alpha}$ is an integer corresponding to the number of publications of author $i$ in subfield $\alpha$. The previous condition can hence by formalised as
\begin{equation}
	 RCA  = \frac{	\frac{w_{i \alpha}}{\sum_{\alpha'}w_{i \alpha'}}     }    {\frac{\sum_{i'}w_{i' \alpha}}{\sum_{i' \alpha'}w_{i' \alpha'}}.    } > 1.
\end{equation}
This filter, known as the Revealed Comparative Advantage (RCA) index, was introduced in 1965 in Ref.~\cite{Balassa1965} and has been used previously to filter bipartite networks, as in Ref.~\cite{hidalgo2009eci}. Differently from other alternatives, it guarantees that each author is active on at least one field. We limit our analysis to authors with at least $N=5$ publications in our reconstructed physics dataset, in order to drop all the authors whose contribution to physics is marginal. This set covers 135,877 authors.

\begin{figure}[ht]
\centering
\includegraphics[width=0.8\textwidth]{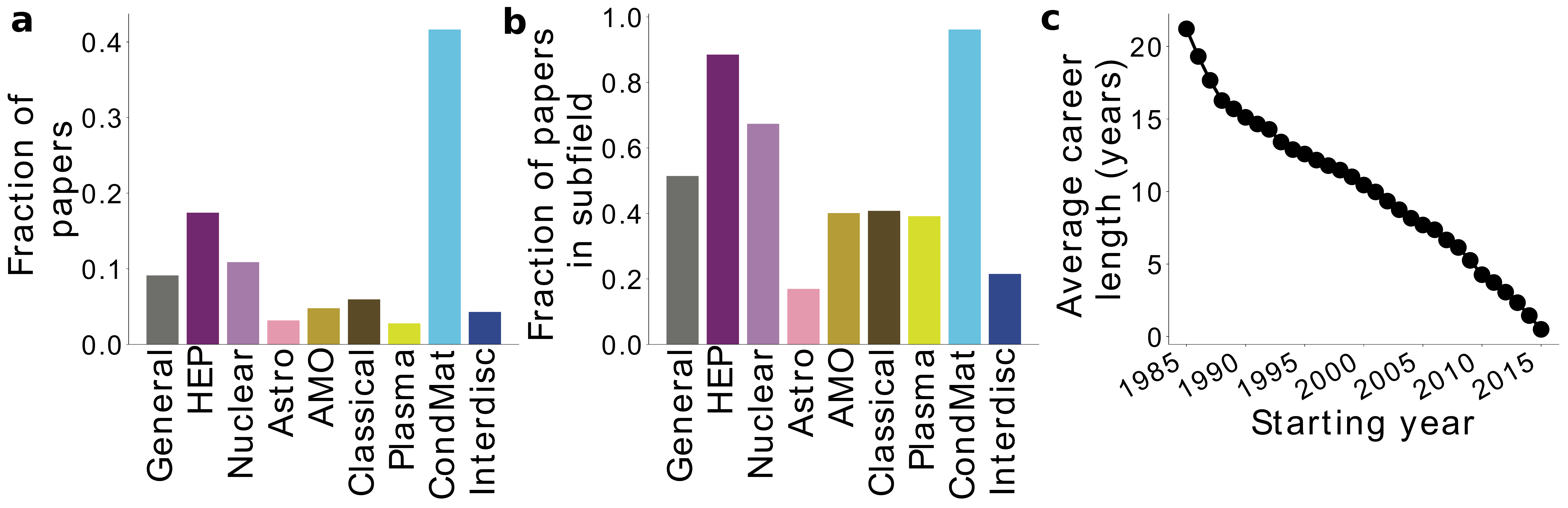}
\caption{\textbf{Basic features of authors in our reconstructed physics dataset.} \textbf{a} Average publication shares across subfields of a physicist. \textbf{b} For authors validated in a subfield, average fraction of publications in that subfield. \textbf{c} Average career length measured in years as a function of the starting year of a career.}
\label{fig:authors}
\end{figure}

The average distribution $w_{i \alpha}$ of subfields per author is shown in Fig.\ref{fig:authors}a. In Fig.\ref{fig:authors}b we show the average fraction of papers in each subfield for authors statistically validated in a given area. This plot is similar to that of Fig.1c of the main text, but reports more fine-grained information about the involvement of physicists in the subfields to which they are assigned. As shown, the share of publication in the subfield of belonging is the highest for authors in {\sf Cond Mat}, {\sf HEP} and {\sf Nuclear}. Last, in Fig.\ref{fig:authors}c  we report the average career length measured in years, of physicists starting publishing in a given year. As expected, the earlier the starting year, the longer the average time span between the first and last publications of a physicist.

\begin{figure}[ht]
\centering
\includegraphics[width=0.4\textwidth]{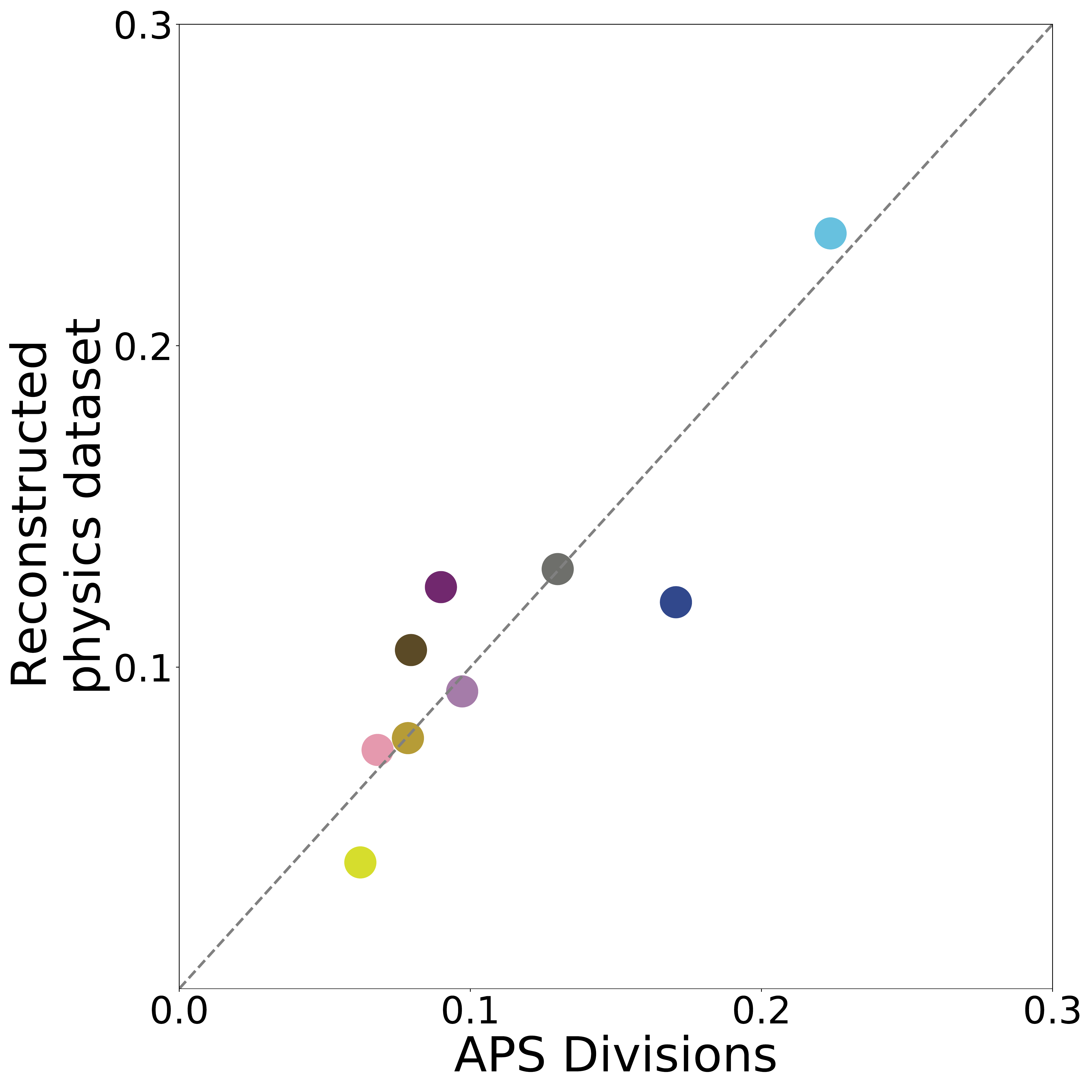}
\caption{\textbf{Comparison of the fraction of physicists associated to the different subfields and the members of the APS Divisions.} Correlation between the two distributions is high, with a cosine similarity of 0.98.}
\label{fig:WoS_scatter_member}
\end{figure}

{\bf Validation:} To test the robustness of our subfield categorisation at the author level, we compared the numbers of authors working in each subfield with the number of APS members registered across APS Divisions~\cite{APSDivisions}. In Fig.~\ref{fig:WoS_scatter_member} we report the scatterplot between the two datasets, with a cosine similarity of 0.98. The full mappings between the APS Divisions and our subfield scheme is reported in Table~\ref{tab:comparison_authors}.

\begin{table}
\begin{center}
 \begin{tabular}{||c | c||} 
 \hline
 \textbf{Subfield} & \textbf{APS Divisions} \\ 
 \hline\hline
  {\sf General} & Computational Physics, Quantum Information, Gravitation\\ 
 \hline
  {\sf HEP} &  Particles \& Fields \\
 \hline
  {\sf Nuclear} & Nuclear Physics, Physics of Beams \\
   \hline
  {\sf Astro} & Astrophysics\\
   \hline
  {\sf AMO} & Atomic, Molecular \& Optical \\
   \hline
  {\sf Classical} & Fluid Dynamics \\
   \hline
  {\sf Plasma} & Plasma Physics\\
   \hline
  {\sf CondMat} & Condensed Matter Physics, Laser Science, Polymer Physics \\
  \hline
  {\sf Interdisc} & Biological Physics, Materials Physics, Chemical Physics\\
 \hline
\end{tabular}

\end{center}
\caption{\textbf{Mapping of physics categories from the APS Divisions into the physics subfield scheme.}}
\label{tab:comparison_authors}
\end{table}

\section{Author disambiguation}
\label{sectionSI:disambiguation}

\begin{figure}[ht]
\centering
\includegraphics[width=0.4\textwidth]{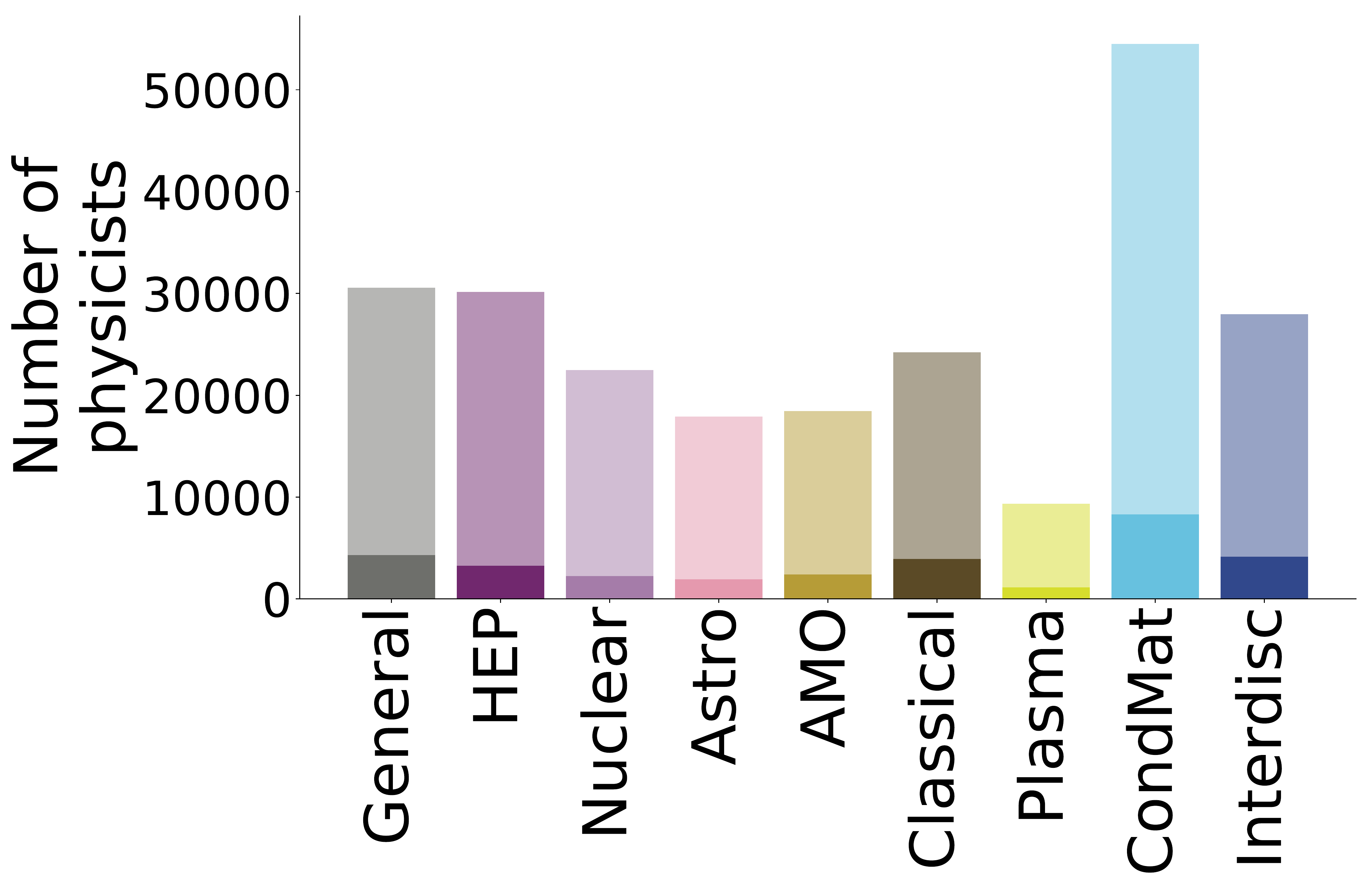}
\caption{\textbf{Testing author disambiguation.} Number of authors working in each subfield: plain color (reduced set of 15,982 authors difficult to disambiguate), faded color (all other physicists). The cosine similarity between the distribution across subfields of the full set of physicists, and the set without authors hard to disambiguate, is 0.99.}
\label{fig:disambiguation}
\end{figure}

A common problem in the analysis of scientific careers is that of author disambiguation~\cite{smalheiser2009}. Our census of physics is based on merging paper information on subfield and author information on publications provided by the WoS. Our analysis has been undertaken on the latest available version of WoS which, differently from the previous one, has a built-in author disambiguation, where authors are not classified by a name but by a specific author ID. A single author ID is associated to a unique author, and can be associated to several author names when the publications authored by the same individual report slightly different name formats. Similarly, two homonyms, but distinct individuals  with the same author name are associated to different author IDs.  Nevertheless, we are aware that a perfect disambiguation is a goal which is impossible to achieve. For such a reason, we decided to test the robustness of our results by replicating the analysis reported in the main text after excluding a subset of authors with names which are known to be particularly hard to disambiguate. In particular, we focused on the most common 100 Chinese and 200 Korean names~\cite{chinesenames,koreannames}, which correspond to 504,538 distinct author IDs in the WoS dataset, 15,982 of which are present also in our subset of physicists. Overall, results were shown to be extremely robust to the elimination of such authors. As an example, we report in Fig.\ref{fig:disambiguation} the starting point of our analysis, i.e. the authors distribution across subfields. The cosine similarity between the distribution across subfields of the full set and the reduced set of physicists, without authors difficult to disambiguate, is 0.99.

It is worth to mention that highly curated data-repositories with very good author disambiguation is available for some subfields. For instance, the well-known HEP-INSPIRE dataset has an extremely valid author disambiguation, especially needed for fields where most publications are done by large collaborations. However, it is difficult to map the HEP-INSPIRE author disambiguation into the built-in WoS author disambiguation. On top of this, we believe that such merge would not add validity to our analysis, as conversely would introduce a bias into the dataset, where authors publishing in different subfields are classified according to different disambiguation procedures.

\clearpage

\section{Null-models for co-activities and transitions between subfields}
\label{sectionSI:nullmodels}
In Fig.1d we map the relation between physics subfields into a network, where nodes represent subfields, and weighted links describe significant co-activity between them. Let us consider a set of $N$ physicists, and two subfields $\alpha$ and $\beta$ with respectively $N^{\alpha}$ and $N^{\beta}$ physicists. We define the co-activity $C^{\alpha \beta}$ between the two subfields as the ratio between the number of physicists $N^{\alpha \beta}$ working on both subfields $\alpha$ and $\beta$, and the expected number $\hat N^{\alpha \beta} = (N^{\alpha}   N^{\beta} )/N $. Starting from the link with the highest weight, we plot the minimum number of links needed to have a connected network. All reported links have $C>1$, meaning that only edges with co-activity higher than what expected at random (given the size of the subfields) are shown.

In Fig.2b we show flows of physicists from the subfield(s) of their first publication, to the subfield(s) where their activity is significant (RCA>1). Let us consider the number of physicists $F^{\alpha | \beta}$ working in subfield $\alpha$ who started their career by publishing in subfield $\beta$, so that $\sum_{\beta} F^{\alpha | \beta}=N^{\alpha}$. Subfield $\beta$ is significantly contributing to subfield $\alpha$ only if $F^{\alpha | \beta} / N^{\alpha}$ is greater than the total fraction of physicists whose first publication is in subfield $\beta$ (reported in the rectangles on the top). Only significant flows are shown.

\section{LHC and the HEP 2010 peak}
\label{sectionSI:LHC}

\begin{figure}[ht]
\centering
\includegraphics[width=0.4\textwidth]{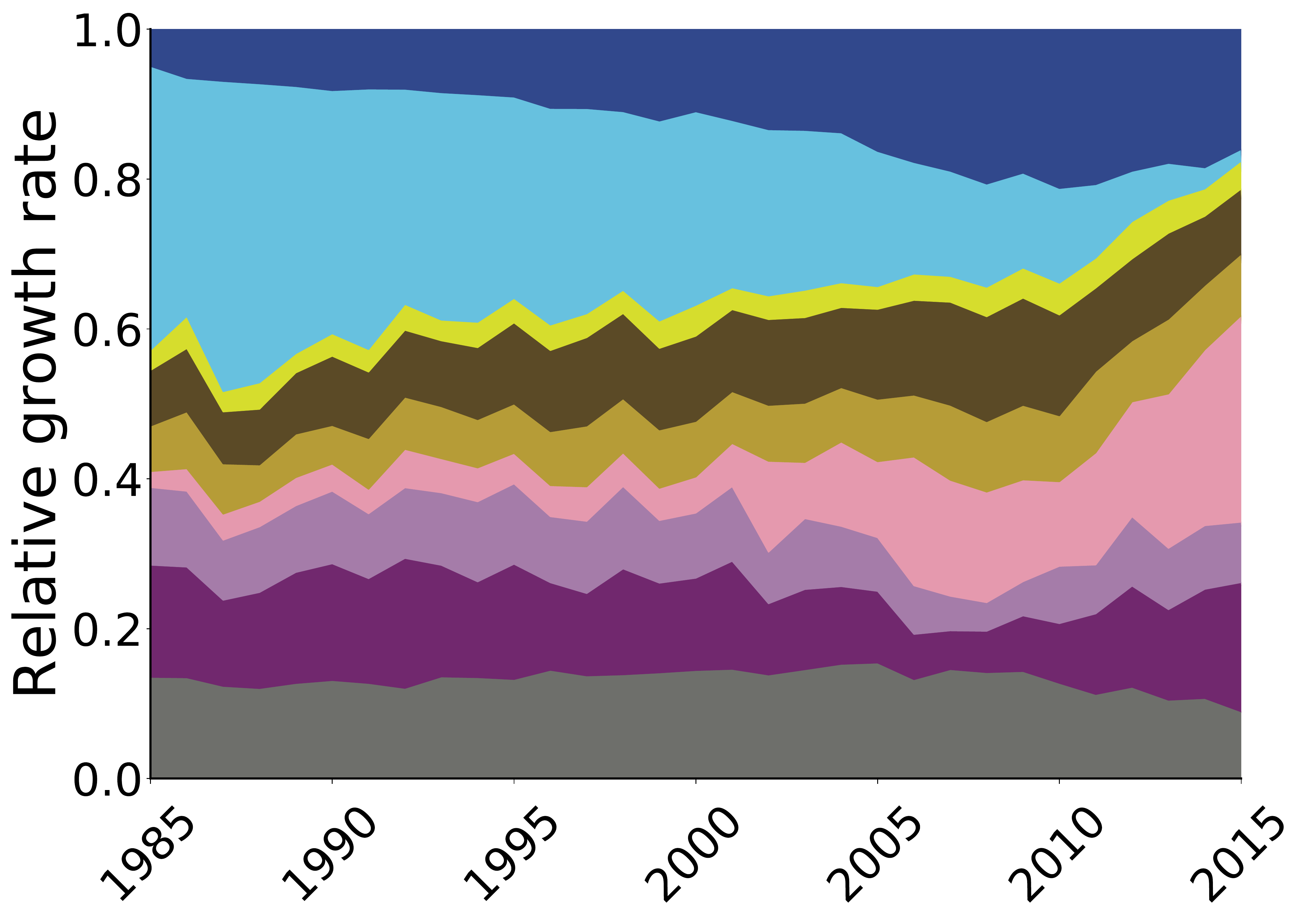}
\caption{\textbf{Relative growth rate of subfields after removing new 2010 HEP authors connected to the activities of LHC.} No peak is observed for {\sf HEP} authors in 2010.}
\label{fig:LHC}
\end{figure}

In Fig.2a we show over the years the relative number of new authors entering each subfield. We notice that {\sf HEP} is characterised by a large peak in 2010. For this reason we looked at all the first publications of new {\sf HEP} authors in 2010, and searched for the collaborations responsible for each paper. We found that $76\%$ of the new {\sf HEP} authors in 2010  have a first publication which is connected to the opening of LHC, either directly through the ATLAS, CMS and LHCb collaborations.~\cite{ATLASCMS2010}, or indirectly (Ref.~\cite{ALICE2010} of the ALICE collaboration takes advantage of results by LHC). 
These new authors also amount to the $21\%$ of the total number of new physicists across subfields, explaining the observed peak for {\sf HEP}. In Fig.~\ref{fig:LHC} we show the yearly fraction of physicists who published their first paper in a new subfield, after removing all new 2010 {\sf HEP} authors connected to the activities of LHC. As displayed, the peak at 2010 for {\sf HEP} disappears.

\section{Chaperone effect}
\label{sectionSI:chaperone}

In Fig.3c we computed the number of chaperoned authors across subfields. The Chaperone effect was originally investigated in Ref.~\cite{Sekara2018} for scientific venues, measured in terms of scientists making the transition from non-last to last (senior / PI) authors in papers published in a journal. Here, as we are interested in the relations, as well as migration between physics subfields, we focused on a simplified version of such chaperone measure $c$ , computing the fraction of physicists first publishing in a subfield who have as co-authors at least one scientist who has already published in the  area.

\begin{figure}[ht]
\centering
\includegraphics[width=0.4\textwidth]{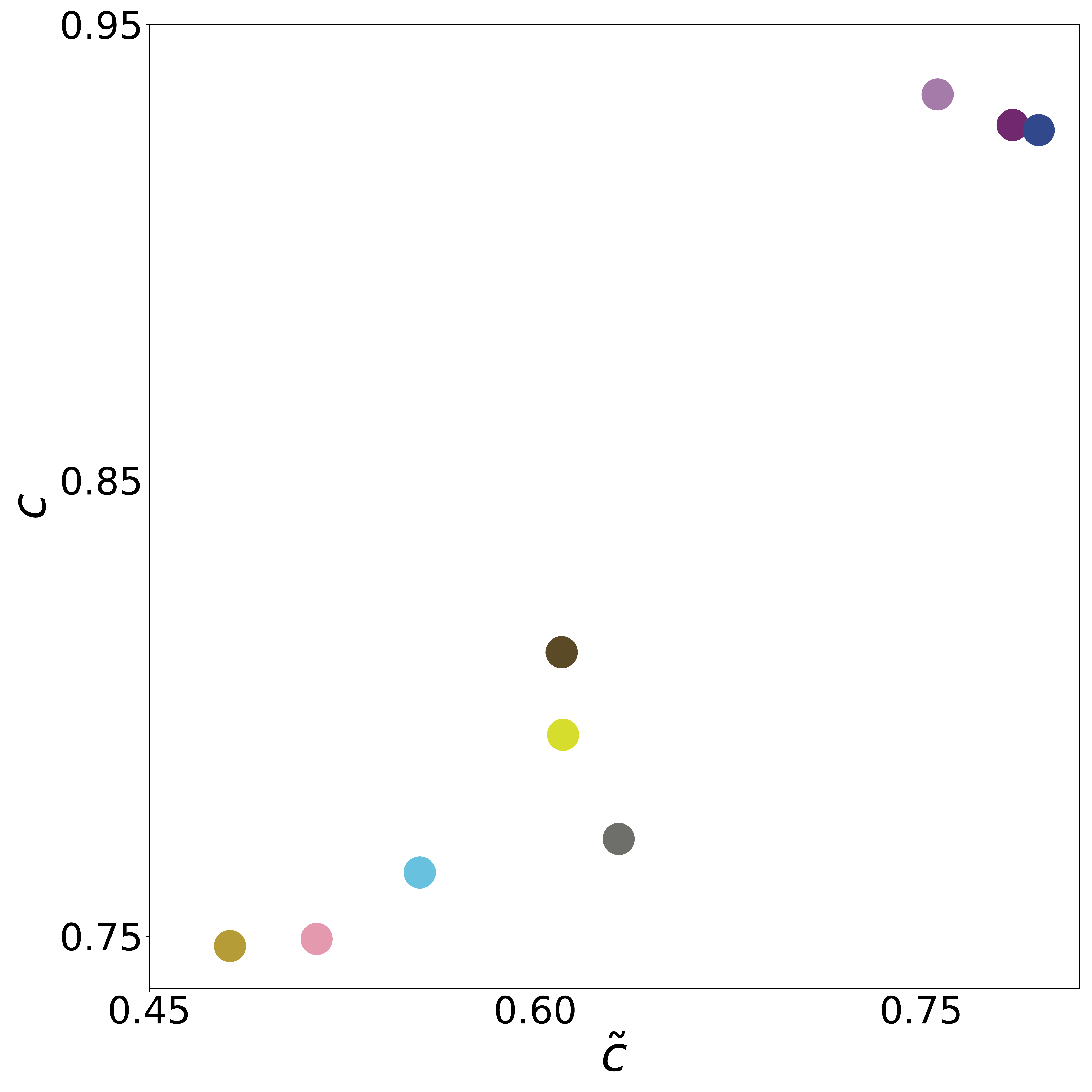}
\caption{\textbf{Comparison between two measures of Chaperone effect.} Scatterplot between the original measure $c$ to quantify the number of chaperoned authors, and the fractional measure $\tilde c$. The values of two variables across subfields in our dataset are highly correlated.}
\label{fig:chap}
\end{figure}

Despite being intuitive and close to the variable used in Ref.~\cite{Sekara2018}, this measure might not prove adequate in the case of subfields characterised by publication through large-scale collaborations. For such a reason, we tested our results against $\tilde c$, a variant of the chaperone index. Given the first publication of a scientists in a subfield, $\tilde c$ measures the average fraction of co-authors who have already published in the area. As shown in Fig.\ref{fig:chap}, in the case of our data $c$ and $\tilde c$ are very highly correlated, with a cosine similarity of $0.99$.

\clearpage

\section{Authors impact and citation rates across subfields}
\label{sectionSI:citations}
Top authors across subfields have very different impact, as shown in Fig.3g. This is mainly a consequence of different productivities, rather than diverse citation patterns across subfields. Indeed, the typical number of papers produced by top authors is very heterogenous across physics communities (Fig.3f). In contrast, we found  that the number of citations per paper is rather constant across subfields: the average is $27.3$, with all subfields falling within $1.8$ standard deviation from this value. For example, papers published in {\sf HEP} and {\sf Interdisc} receive on average respectively $27.4$ and $33.9$ citations, despite the much larger impact of {\sf HEP} authors. Similar results are obtained for the medians of paper citations across subfields. The average median across physics communities is 9.0, the standard deviation of the median across subfields is 1.1, and all subfields are at most 1.7 standard deviation away from the global median. The median of paper citations for {\sf HEP} and {\sf Interdisc} are respectively 9 and 11.

\section{The physics Nobel prizes}
\label{sectionSI:nobel}
In Fig.3j we show the distribution of Nobel prizes awarded in physics across subfields. Data on Nobel prizes in physics are available on the Nobel prize website~\cite{nobelphysicslink}.
We report all awards since 1985 in order to be consistent with the rest of our data-driven analysis of careers in physics. All such awards are accompanied by a motivation which allows to assign the crucial discovery or stream of research that led to the Nobel prize to one or more physics subfields. In the considered time span (1985-2017), 82 scientists were awarded the Nobel prize in physics.
\clearpage

\bibliography{success_natrevphys_si.bib}